\documentclass[psfig]{aa}
\usepackage{graphics}
\usepackage{psfig}

\begin{document}


\title {Chromospheric activity, lithium and radial velocities
of single late-type stars possible members of young moving groups
\thanks{Based on observations made
with the 2.2m telescope of the German-Spanish Astronomical Centre, 
Calar Alto (Almer\'{\i}a, Spain), 
operated by the Max-Planck-Institute for Astronomy,
Heidelberg, jointly with the Spanish National Commission for Astronomy,  
with the Nordic Optical Telescope (NOT),
 operated on the island of La Palma jointly by Denmark, Finland,
 Iceland, Norway and Sweden, in the Spanish Observatorio del 
 Roque de Los Muchachos of the Instituto de Astrof\'{\i}sica de Canarias,
and with the Isaac Newton Telescope (INT)
 operated on the island of La Palma by the Isaac Newton Group in
 the Spanish Observatorio del Roque de Los Muchachos of the
 Instituto de Astrof\'{\i}sica de Canarias.
} 
\thanks{Tables 1, 4, 5 also available in electronic form}
}

\titlerunning{Spectroscopic analysis of 
single late-type stars members of young moving groups}

\author{
D.~Montes
\and J. L\'opez-Santiago
\and M.J.~Fern\'{a}ndez-Figueroa
\and M.C.~G\'alvez
}

\offprints{ D.~Montes}
\mail{dmg@astrax.fis.ucm.es}

\institute{
Departamento de Astrof\'{\i}sica,
Facultad de Ciencias F\'{\i}sicas,
 Universidad Complutense de Madrid, E-28040 Madrid, Spain\\
E-mail: dmg@astrax.fis.ucm.es
}

\date{Received 23 July 2001 / Accepted 26 September 2001}

\abstract{
We present here high resolution echelle spectra 
taken during three observing runs 
of 14 single late-type stars
identified in our previous studies (Montes et al. 2001b, hereafter Paper~I)
as possible members of different young stellar kinematic groups
(Local Association (20 - 150 Myr),
Ursa Major group (300 Myr),
 Hyades supercluster (600 Myr), and
 IC 2391 supercluster (35 Myr)). 
Radial velocities have been determined by cross correlation
with radial velocity standard stars and used together with precise
 measurements of proper motions and parallaxes taken from
Hipparcos and Tycho-2 Catalogues,
to calculate Galactic space motions ($U$, $V$, $W$) and to apply 
Eggen's kinematic criteria.
The chromospheric activity level of these stars have been  analysed
using the information provided for several optical spectroscopic
features (from the Ca~{\sc ii} H \& K to Ca~{\sc ii} IRT lines)
that are formed at different heights in the chromosphere.
The Li~{\sc i} $\lambda$6707.8~\AA$\ $ line equivalent width (EW) 
has been determined and compared in the $EW$(Li~{\sc i}) versus 
spectral type diagram with the $EW$(Li~{\sc i}) of stars members
of well-known young open clusters of different ages,
in order to obtain an age estimation.
All these data allow us to analyse in more detail
the membership of these
stars in the different young stellar kinematic groups.
Using both, kinematic and spectroscopic criteria 
we have confirmed 
PW And, V368 Cep, V383 Lac, EP Eri, DX Leo, HD 77407, and EK Dra 
as members of the Local Association and
V834 Tau, $\pi^{1}$ UMa, and GJ 503.2 
as members of the Ursa Major group.
A clear rotation-activity dependence has been found in these stars.
}
 
\maketitle

\section{Introduction}

It has long been known that in the solar vicinity there are several
kinematic groups of stars that share
the space motions of well-known open clusters.
Eggen (1994) defined a "supercluster" (SC) as a group of stars,
gravitationally unbound, that share the same kinematics and may
occupy extended regions in the Galaxy, and a "moving group" (MG)
as the part of the supercluster that enters the solar neighbourhood
and can be observed  all over the sky.
The origin of these stellar kinematic groups (SKG)
could be the evaporation of an open cluster,
the remnants of a star formation region
or a juxtaposition of several little star formation bursts
at different epochs in adjacent cells of the velocity field.
The youngest and best-documented SKG are:
the Hyades supercluster (600 Myr)
the Ursa Major group (Sirius supercluster) (300 Myr),
the Local Association or Pleiades moving group (20 to 150 Myr),
the IC 2391 supercluster (35-55 Myr), and
the Castor moving group (200 Myr)
(see Montes et al. 2001b, hereafter Paper~I, and references therein).

Well-known members of these SKG are mainly early-type stars
and few studies have been centered on late-type stars.
However, the identification of a significant number of 
late-type population in these young SKG
is extremely important for the study of the
chromospheric activity and could lead to a better understanding of
star formation history in the solar neighbourhood.
In our previous work (Montes et al. 2000a, 2001a; Paper~I) 
a sample of late-type stars
of previously established members and possible new candidates
to these five young SKG have been identified.
In order to better establish the membership
of these candidate stars in the different young SKG,
we have started a program of high resolution
echelle spectroscopic observations. 
The spectroscopic analysis of these stars allows us to obtain
a better determination of their radial velocity, 
lithium ($\lambda$6707.8 line) equivalent width , rotational velocity 
and the level of chromospheric activity.
We will use all these new data to study in detail 
the kinematics (Galactic space motions ($U$, $V$, $W$)) of these stars,
apply age-dating methods for late-type stars, and in this way 
analyse in more detail the membership 
of these stars in the different SKG.

We present here the results of our first spectroscopy studies
of a sample of 14 single late-type stars selected by us in Paper~I as
young disk stars or possible members of some of the above mentioned 
young SKG.
The high resolution echelle spectra analysed here  
were taken during three observing runs (from 1999 to 2000)
and include all the optical chromospheric activity indicators
from the Ca~{\sc ii} H \& K to Ca~{\sc ii} IRT lines 
as well as the Li~{\sc i} $\lambda$6707.8 line.

In Sect.~2 we give the details of our observations and data reduction.
The radial velocity and Galactic space-velocity components ($U$, $V$, $W$)
determination is described in Sect.~3.
The Li~{\sc i} $\lambda$6707.8 line is analysed in Sect.~4.
The different chromospheric activity indicators are analysed in
Sect.~5.
Individual results for each star are reported in Sect.~6.
Finally, in Sect.~7 the discussion and conclusions are given.

\begin{figure*}
{\psfig{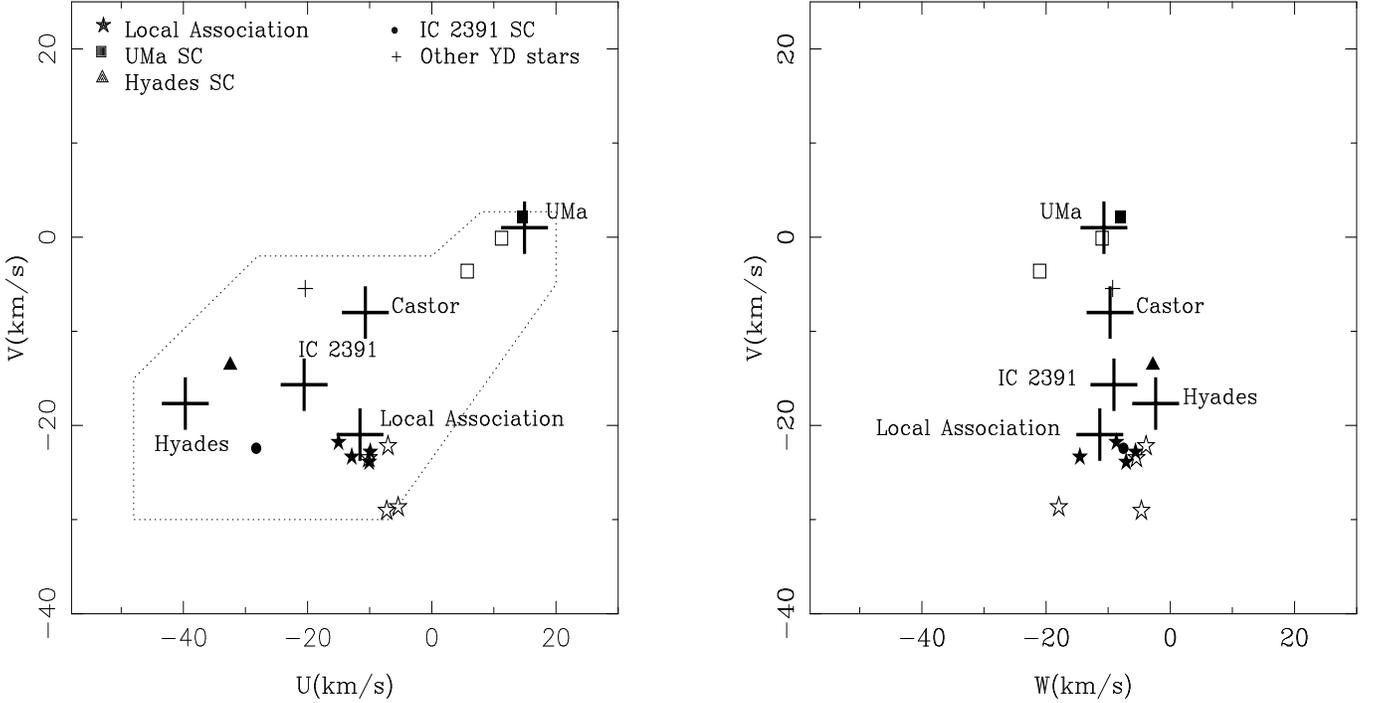}}
\caption[ ]{($U$, $V$) and ($W$, $V$) planes (Boettlinger Diagram) for our star sample.
We plot with different symbols the stars belonging
to the different stellar kinematic groups.
Filled symbols are stars that satisfied both of Eggen's criteria
(peculiar velocity, $PV$, and radial velocity, $\rho_{c}$),
open symbols are other possible members.
Big crosses are plotted in the central position of each group.
The dashed line represents the  boundaries
that determine the young disk population as defined by Eggen (1984a, 1989).
\label{fig:uvw}}
\end{figure*}

\section{Observations and Data Reduction}
 
The spectroscopic echelle observations of the stars
analysed in this paper were obtained during three observing runs:

{1)} {2.2m-FOCES 1999/07} \\
This took place on 
24 -29 July 1999 using the 2.2~m telescope at the German Spanish Astronomical
Observatory (CAHA) (Almer\'{\i}a, Spain).
The Fibre Optics Cassegrain Echelle Spectrograph (FOCES)
(Pfeiffer et al. 1998)
was used with a 2048$^{2}$ 15$\mu$ LORAL$\#$11i CCD detector.
The wavelength range covers from
3910 to 9075 \AA$\ $ in 84 orders.
The reciprocal dispersion ranges from 0.03 to 0.07 \AA/pixel
and the spectral resolution,
determined as the full width at half maximum (FWHM)
of the arc comparison lines, ranges from 0.09 to 0.26 \AA.

{2)} {NOT-SOFIN 1999/11} \\
Observations taken on  26 -27 November 1999
using the 2.56~m Nordic Optical Telescope (NOT) located
at the Observatorio del Roque de Los Muchachos (La Palma, Spain).
The Soviet Finnish High Resolution Echelle Spectrograph
(SOFIN) was used with an echelle grating (79 grooves/mm),
 Astromed-3200 camera and a 1152$\times$770
pixel EEV P88200 CCD detector. The wavelength range covers from
3525 to 10425 \AA$\ $ in 44 orders.
The reciprocal dispersion ranges from 0.06 to 0.17 \AA/pixel
and the spectral resolution (FWHM) from 0.14 to 0.32 \AA.

{3)} {INT-MUSICOS 2000/01} \\
Observations make on 18-22 January 2000
with the 2.5~m Isaac Newton Telescope (INT)
at the Observatorio del Roque de Los Muchachos (La Palma, Spain)
using the {ESA-MUSICOS} spectrograph.
This is a fibre-fed cross-dispersed echelle spectrograph,
built as a replica of the
first MUSICOS spectrograph
(Baudrand \& B\"ohm 1992) and developed as part of
MUlti-SIte COntinuous Spectroscopy 
(MUSICOS\footnote{http://www.ucm.es/info/Astrof/MUSICOS.html}) project.
During this observing run,
a 1024$^{2}$ 24$\mu$ TEK5 CCD detector was used,
obtaining wavelength coverage from 4430~\AA$\ $ to 10225~\AA$\ $ in 73 orders.
The reciprocal dispersion ranges from 0.07 to 0.15 \AA$\ $
and the spectral resolution (FWHM) from~0.16 to 0.30 \AA.

The sample of late-type stars analysed in this paper as well as the
the non-active stars used as reference stars in the spectral subtraction
and the radial velocity standards used in the radial velocity determinations
are listed in Tables \ref{tab:obslog} and \ref{tab:par}.
In Table~\ref{tab:obslog} we give the observing log.
For each observation we list date, UT
and the signal to noise ratio ($S/N$) obtained in the Ca~{\sc ii} H \& K and 
H$\alpha$ line regions.
Table~\ref{tab:par} shows the name, HD number, and
other stellar parameters such as the spectral type (T$_{\rm sp}$),
color indexes $V$--$R$ and $B$--$V$, rotational velocity ($v\sin{i}$), 
rotational period ($P_{\rm phot}$). 
The $V$--$R$ and $B$--$V$ color indexes are obtained from the
relation with spectral type given by Landolt-B\"{o}rnstein (Schmidt-Kaler 1982)
when individual values are not given in the literature.
Other parameters are taken from the references given
in the individual results of each star.
The Galactic space motions ($U$, $V$, $W$) given in Table~\ref{tab:par} 
have been determined by us as explained in the next section.

The spectra have been extracted using the standard
reduction procedures in the
IRAF\footnote{IRAF is distributed by the National Optical Observatory,
which is operated by the Association of Universities for Research in
Astronomy, Inc., under contract with the National Science Foundation.}
 package (bias subtraction,
flat-field division and optimal extraction of the spectra).
The wavelength calibration was obtained by taking
spectra of a Th-Ar lamp.
Finally, the spectra were normalized by
a polynomial fit to the observed continuum.

\begin{table*}
\caption[]{Observing log
\label{tab:obslog}}
\begin{flushleft}
\scriptsize
\begin{tabular}{lcccccccccccccc}
\noalign{\smallskip}
\hline
\noalign{\smallskip}
Name &
\multicolumn{4}{c}{2.2m-FOCES 1999/07} &\ &
\multicolumn{4}{c}{NOT-SOFIN 1999/11} &\ &
\multicolumn{4}{c}{INT-MUSICOS 2000/01} \\
\cline{2-5}\cline{7-10}\cline{12-15}
\noalign{\smallskip}
     &
\tiny Day  & \tiny UT &  \tiny $S/N$ & \tiny $S/N$ &\ &
\tiny Day  & \tiny UT &  \tiny $S/N$ & \tiny $S/N$ &\ &
\tiny Day  & \tiny UT &  \tiny $S/N$ & \tiny $S/N$ 
\\
     &
 &  &  \tiny (H\&K) & \tiny (H$\alpha$) &\ &
 &  &  \tiny (H\&K) & \tiny (H$\alpha$) &\ &
 &  &  \tiny (H\&K) & \tiny (H$\alpha$)
\scriptsize
\\
\noalign{\smallskip}
\hline
\noalign{\smallskip}
{\bf LA} \\
\noalign{\smallskip}
\hline
\noalign{\smallskip}
PW And   & 25 & 04:09 & 23 & 123 & & 26 & 20:24 & 25 & 107 & &    &     &  &     \\
 "       & 26 & 00:57 & 19 & 110 & & 27 & 21:29 & 20 & 155 & &    &     &  &     \\
 "       & 27 & 02:25 & 13 &  85 & &    &       & &     & &    &     &  &     \\
 "       & 28 & 00:57 & 15 &  76 & &    &       & &     & &    &     &  &     \\
 "       & 29 & 03:01 & 21 & 122 & &    &       & &     & &    &     &  &     \\
 "       & 30 & 01:57 & 20 & 130 & &    &       & &     & &    &     &  &     \\
V368 Cep & 25 & 01:00 & 31 & 168 & & 27 & 20:53 & 13 & 132 & &    &     &  &     \\
 "       & 26 & 00:12 & 16 & 167 & &    &       & &     & &    &     &  &     \\
 "       & 27 & 04:08 & 11 &  80 & &    &       & &     & &    &     &  &     \\
 "       & 28 & 02:14 & 26 & 179 & &    &       & &     & &    &     &  &     \\
 "       & 29 & 00:50 & 16 & 188 & &    &       & &     & &    &     &  &     \\
 "       & 30 & 00:42 & 30 & 224 & &    &       & &     & &    &     &  &     \\
V383 Lac & 25 & 00:20 & 28 & 130 & &    &       & &     & &    &     &  &     \\
         & 25 & 03:45 & 26 & 128 & &    &       & &     & &    &     &  &     \\
 "       & 25 & 23:48 & 11 &  96 & &    &       & &     & &    &     &  &     \\
 "       & 27 & 02:02 & 15 &  83 & &    &       & &     & &    &     &  &     \\
 "       & 28 & 00:33 & 16 &  77 & &    &       & &     & &    &     &  &     \\
 "       & 29 & 00:03 & 17 & 120 & &    &       & &     & &    &     &  &     \\
 "       & 29 & 23:58 & 16 & 112 & &    &       & &     & &    &     &  &     \\
EP Eri   &    &       & &     & & 26 & 22:27 & 23 & 151 & &    &       &     \\
 "       &    &       & &     & & 27 & 23:49 & 27 & 231 & &    &       &     \\
DX Leo   &    &       & &     & & 27 & 05:45 & 13 & 168 & & 19 & 04:11 & - & 138 \\
 "       &    &       & &     & & 28 & 06:17 & 37 & 215 & & 21 & 04:58 & - & 139 \\
 "       &    &       & &     & &    &       &    &     & & 23 & 04:00 & - & 100 \\
GJ 211   &    &       & &     & &    &       &    &     & & 23 & 02:16 & - & 119 \\
HD 77407 &    &       & &     & &    &       &    &     & & 21 & 04:36 & - & 171 \\
 "       &    &       & &     & &    &       &    &     & & 23 & 03:43 & - & 102 \\
EK Dra   &    &       & &     & &    &       &    &     & & 20 & 07:10 & - & 121 \\
 "       &    &       & &     & &    &       &    &     & & 23 & 07:16 & - &  61 \\
\noalign{\smallskip}
\hline
\noalign{\smallskip}
{\bf UMa} \\
\noalign{\smallskip}
\hline
\noalign{\smallskip}
V834 Tau &    &       & &     & & 27 & 01:49 & 14 & 132 & & 23 & 01:13 & - &  56 \\
 "       &    &       & &     & & 28 & 02:25 & 18 & 183 & &    &       & - &     \\
$\pi^{1}$ UMa &  &    & &     & & 28 & 06:47 & 29 & 284 & & 21 & 04:30 & - & 110 \\
 "       &    &       & &     & &    &       & &     & & 23 & 03:34 & - &  94 \\
GJ 503.2 &    &       & &     & &    &       & &     & & 23 & 06:14 & - &  68 \\
\noalign{\smallskip}
\hline
\noalign{\smallskip}
{\bf Others} \\
\noalign{\smallskip}
\hline
\noalign{\smallskip}
LQ Hya (LA) & &       & &     & & 27 & 06:12 & 12 & 104 & & 21 & 04:01 & - & 123 \\
 "       &    &       & &     & & 28 & 07:03 & 15 & 164 & & 23 & 02:29 & - & 108 \\
GJ 577 (HS)&25& 20:19 & 17 &  89 & &    &       & &     & &    &     &  &     \\
 "         &27& 20:16 & 18 &  78 & &    &       & &     & &    &     &  &     \\
 "         &29& 19:57 & 26 & 125 & &    &       & &     & &    &     &  &     \\
GJ 3706
(IC 2391)   & &       & &     & &    &       & &     & & 21 & 06:35 & - & 110 \\
%
\noalign{\smallskip}
\hline
{\bf\tiny Ref. Stars} \\
\noalign{\smallskip}
\hline
\noalign{\smallskip}
107 Psc   & 27 & 04:29 & 41 & 212  \\
GJ 706    & 25 & 00:44 & 37 & 263 \\
 "        & 25 & 20:07 & 35 & 176 \\
GJ 758 *  & 26 & 00:33 & 43 & 195 \\
HR 8088   & 26 & 00:45 & 43 & 226 \\
GJ 639    & 28 & 22:26 & 29 & 148 \\
GJ 679    & 25 & 22:00 & 39 & 156 \\
$\beta$ Oph * & 24 & 23:48 & 31 & 182 \\
 "            & 25 & 19:51 & 25 & 222 \\
 "            & 25 & 19:55 & 25 & 268 \\
 "            & 27 & 20:02 & 22 & 164 \\
 "            & 28 & 19:57 & 39 & 332 \\
 "            & 28 & 20:01 & 42 & 327 \\
 "            & 29 & 19:46 & 32 & 314 \\
61 Cyg A    & 30 & 00:22 & 43 & 367 \\
61 Cyg B    & 30 & 00:31 & 32 & 336 \\
HR 7949   & 26 & 04:39 & 23 & 235 & & 26 & 20:16 & 42 & 180 \\
HR 166    * & 29 & 04:37 & 51 & 263 & & 26 & 22:50 & 31 & 102 & & 22 & 20:17 & - & 71 \\
 "          &    &       &    &     & & 27 & 22:52 & 43 & 162 & & \\
HR 222    * &    &       &    &     & & 26 & 23:06 & 27 & 119 \\
HR 8832     &    &       &    &     & & 27 & 20:29 & 36 & 228 \\
$\beta$ Gem * &  &       &    &     & &    &       &     &    & & 19 & 01:25 & - & 283 \\
 "            &  &       &    &     & &    &       &     &    & & 21 & 03:35 & - & 149 \\
Sun           &  &       &    &     & &    &       &     &    & & 23 & 07:53 & - & 268 \\

%
\noalign{\smallskip}
\hline
\end{tabular}

\end{flushleft}
\end{table*}

\begin{table*}
\caption[]{Stellar parameters 
\label{tab:par}}
\begin{flushleft}
\scriptsize
\begin{tabular}{l l c c c c c r r r r }
\noalign{\smallskip}
\hline
\noalign{\smallskip}
Name & HD/BD& {T$_{\rm sp}$}& {$V$--$R$}& {$B$--$V$}& 
$v\sin{i}$ &{$P_{\rm phot}$}&
$U\pm \sigma_{U}$ & $V \pm \sigma_{V}$ & $W \pm \sigma_{W}$ \\
     &      &               &       &       &(km s$^{-1}$)&   (days)         &
(km s$^{-1}$) & (km s$^{-1}$) & (km s$^{-1}$) \\
\noalign{\smallskip}
\hline
\noalign{\smallskip}
{\bf LA} \\
\noalign{\smallskip}
\hline
\noalign{\smallskip}
PW And    & HD 1405    & K2V    & 0.74 & 1.00 & 23.4  & 1.75 & -5.42$\pm$0.33 & -28.69$\pm$0.63 &   -17.94$\pm$0.74 \\
V368 Cep  & HD 220140  & K1V    & 0.61 & 0.87 & 16.1 & 2.74  & -10.16$\pm$0.25 & -23.48$\pm$0.16 &  -5.45$\pm$0.10  \\
V383 Lac  & BD+48 3686 & K1V    & 0.69 & 0.83 & 19.8 & 2.42  & -7.06$\pm$1.43 & -22.19$\pm$0.34 &  -3.90$\pm$0.86 \\
EP Eri    & HD 17925   & K1V    & 0.69 & 0.86 & 6.2  & 6.85  & -15.01$\pm$0.10 & -21.80$\pm$0.18 &  -8.68$\pm$0.11 \\
DX Leo    & HD 82443   & K0V    & 0.64 & 0.78 & 6.2  & 5.377 & -9.91$\pm$0.15 & -22.83$\pm$0.36 &  -5.61$\pm$0.23 \\
GJ 211    & HD 37394   & K1V    & 0.69 & 0.84 & 4.0  & 10.86 & -12.89$\pm$0.23 & -23.35$\pm$0.27 & -14.55$\pm$0.18 \\
HD 77407  & BD+38 1993 & G0     & 0.50 & 0.61 & 7.0* &       & -10.10$\pm$0.30 & -23.91$\pm$0.70 &  -7.12$\pm$0.38 \\
EK Dra    & HD 129333  & G1.5V  & 0.52 & 0.59 & 17.3 & 2.787 & -7.25$\pm$0.32 & -29.07$\pm$0.42 &  -4.65$\pm$0.35 \\
%
\noalign{\smallskip}
\hline
\noalign{\smallskip}
{\bf UMa} \\
\noalign{\smallskip}
\hline
\noalign{\smallskip}
V834 Tau  & HD 29697   & K4V    & 0.91 & 1.09 & 9.5  & 3.936& 5.71$\pm$0.31 &  -3.60$\pm$0.09 & -21.04$\pm$0.37 \\
$\pi^{1}$ UMa & HD 72905& G1.5V & 0.52 & 0.62 & 9.7  & 4.68 & 11.24$\pm$0.09 &  -0.10$\pm$0.10 & -10.99$\pm$0.09   \\
GJ 503.2  & HD 115043  & G2V    & 0.53 & 0.67 & 7.5  &      & 14.52$\pm$0.26 &   2.19$\pm$0.21 &  -8.08$\pm$0.27   \\
\noalign{\smallskip}
\hline
\noalign{\smallskip}
{\bf Others} \\
\noalign{\smallskip}
\hline
\noalign{\smallskip}
LQ Hya    & HD 82558   & K2V   & 0.64 & 0.91 & 25   & 1.66 & -20.36$\pm$0.35 &  -5.45$\pm$0.17  & -9.28$\pm$0.28    \\
GJ 577    & HD 134319  & G5    & 0.54 & 0.68 &      & 4.448 & -32.45$\pm$1.05 &  -13.59$\pm$0.35 &  -2.82$\pm$0.20    \\
GJ 3706   & HD 105631  & K0V   & 0.64 & 0.80 & 4.5  &      & -28.26$\pm$0.86 & -22.43$\pm$0.61  & -7.55$\pm$1.93    \\
%
\noalign{\smallskip}
\hline
\noalign{\smallskip}
{\bf\tiny Ref. Stars} \\
\noalign{\smallskip}
\hline
\noalign{\smallskip}
107 Psc       & HD 10476  & K1V  \\
GJ 706        & HD 166620 & K2V  \\
GJ 758 *      & HD 182488 & G8V \\
HR 8088       & HD 201196 & K2IV \\
GJ 639        & HD 151877 & K7V \\
GJ 679        & HD 159222 & G5V \\
$\beta$ Oph * & HD 161096 & K2III \\
61 Cyg A      & HD 201092 & K5V \\
61 Cyg B      & HD 201092 & K7V \\
HR 7949       & HD 197989 & K0III \\
HR 166 *      & HD 3651   & K0V \\
HR 222 *      & HD 4628   & K2V \\
HR 8832       & HD 219134 & K3/4V  \\
$\beta$ Gem * & HD 62509  & K0III  \\
Sun           & -         & G2V  \\
\hline
\noalign{\smallskip}
\end{tabular}

\end{flushleft}
\end{table*}

\section{Radial Velocities and Space Motions}

Heliocentric radial velocities were determined by using the
 cross-correlation technique.
The spectra of the program stars were cross-correlated order by order,
using the routine {\sc fxcor} in IRAF, against spectra of radial velocity
standards of similar spectral types 
(the stars marked with * in Tables 1 and 2)
taken from Beavers et al. (1979).
The velocity is derived for each order 
from the position of the cross-correlation peak.
Radial velocity  errors for each order are computed by {\sc fxcor} based on the 
fitted peak height and the antisymmetric noise as described by 
Tonry \&  Davis (1979).
In Table~\ref{tab:radvel} we list, for each spectrum, the 
heliocentric radial velocities (V$_{\rm hel}$)
and their associated errors ($\sigma_{\rm V}$)
obtained as weighted means of the individual values deduced for each order.
The orders including chromospheric features and prominent telluric lines
have been excluded when determining the mean velocity.
Finally, we also list in Table~\ref{tab:radvel}
a mean value for each star, obtained as weighted mean 
of the individual values deduced for each spectrum.

We have used these radial velocities together with precise measurements
of proper motions and parallaxes taken from Hipparcos (ESA 1997) 
and Tycho-2 (H$\o$g et al. 2000) Catalogues,
to calculate the Galactic space-velocity components ($U$, $V$, $W$) 
in a right-handed coordinated system (positive in the directions of the
Galactic center, Galactic rotation, and the
North Galactic Pole, respectively).
We have modified the procedures in Johnson \& Soderblom (1987)
to calculate $U$, $V$, $W$, and their associated errors.
The original algorithm (which requires epoch 1950 coordinates) is adapted
here to epoch J2000 coordinates
in the International Celestial Reference System (ICRS).
The uncertainties of the velocity components have been obtained using
the full covariance matrix in order to take into account the
possible correlation between the astrometric parameters.
We have used the correlation coefficients provided by Hipparcos (ESA 1997).
The obtained values are given in Table~\ref{tab:par}
and the ($U$, $V$) and ($W$, $V$) planes
(Boettlinger Diagram) are plotted in Fig.~\ref{fig:uvw}.
In this figure we have also plotted (big crosses) the central position given 
in the literature (see Paper~I) of the five young SKG
analysed in this work.

As in Paper~I we have used as membership criteria the position of the stars
in the velocity space
($U$, $V$, $W$) and the Eggen's kinematic criteria 
of deviation of the space motion
of the star from the convergent point (peculiar velocity, $PV$)
and comparison between the observed and calculated ($\rho_{c}$) 
radial velocities.
Using these criteria we have considered 
PW And, V368 Cep, V383 Lac, EP Eri, DX Leo, GJ 211, HD 77407, and EK Dra 
as possible members of the Local Association (LA);
V834 Tau, $\pi^{1}$ UMa, and GJ 503.2
as possible members of the Ursa Major group (UMa);
GJ 577 as a possible member of the Hyades supercluster (HS);
GJ 3706 as a possible member of IC 2391 supercluster; and
LQ Hya as another young disk (YD) star.
More details about the membership of these stars are given in the individual
results of each star.

\begin{figure}
{\psfig{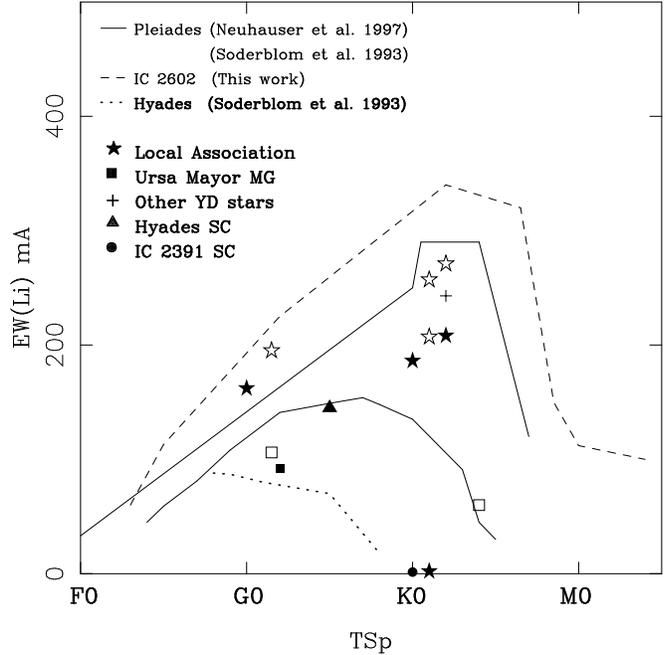}}
\caption[ ]{Li~{\sc i} $\lambda$6707.8 line equivalent width, $EW$(Li~{\sc i})
versus spectral type for our star sample.
Symbols are as in Fig.~\ref{fig:uvw}.
Dashed line represent the upper envelope of $EW$(Li~{\sc i}) observed in the
young open cluster IC 2602; solid lines are the upper and lower envelopes of
 $EW$(Li~{\sc i}) in the Pleiades and the dotted line in the Hyades.
\label{fig:pli}}
\end{figure}

\begin{figure}
{\psfig{figure=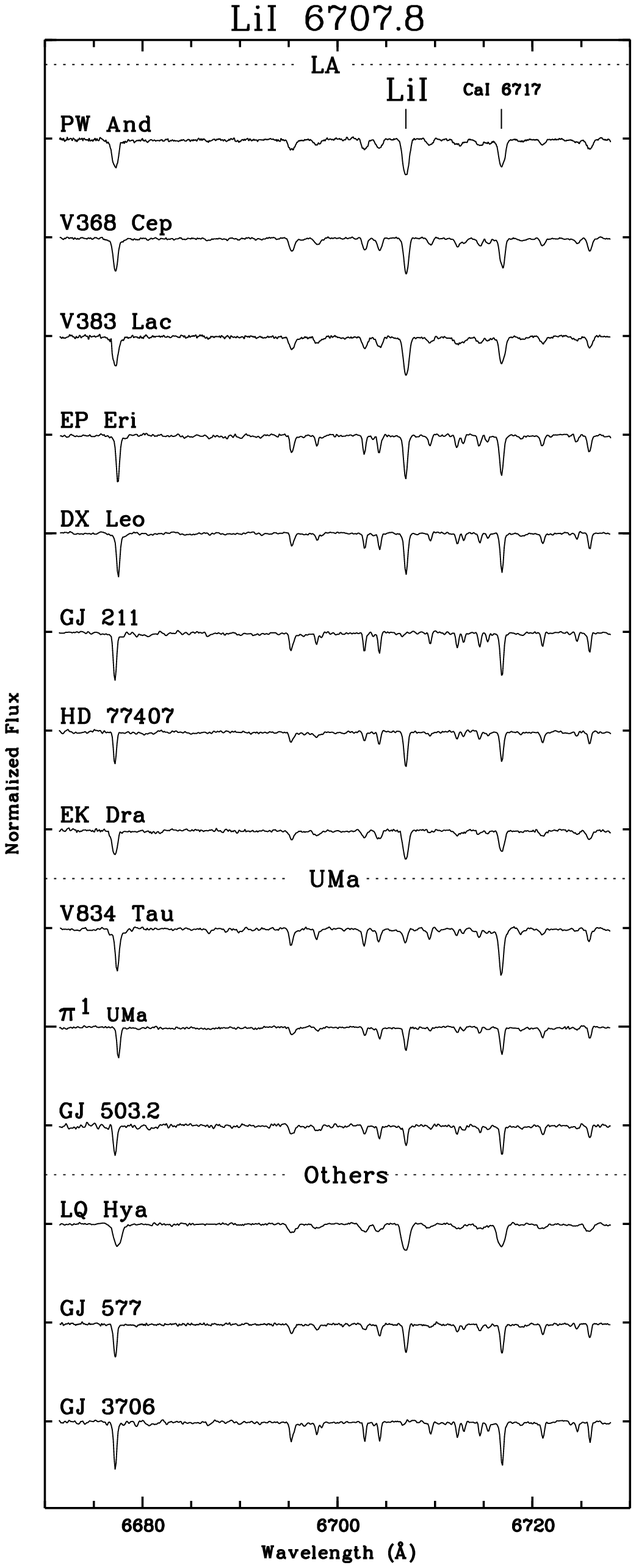,bbllx=33pt,bblly=33pt,bburx=335pt,bbury=777pt,height=18.8cm,width=8.7cm,clip=}}
\caption[ ]{Spectra in the Li~{\sc i} $\lambda$6707.8 line region for
our star sample
\label{fig:li} }
\end{figure}

\section{The Li~{\sc i} $\lambda$6707.8 line}

The resonance doublet of  Li~{\sc i} at $\lambda$6707.8 \AA\
is an important diagnostic of age in late-type stars
since it is destroyed easily by thermonuclear reactions in the
stellar interior.
This line is included in
our echelle spectra in all the observing runs.
At this spectral resolution and with the rotational velocity
($v\sin{i}$ $>$ 8 km s$^{-1}$) of the observed stars
the Li~{\sc i} line is blended with
the nearby Fe~{\sc i} $\lambda$6707.41~\AA\ line.
We have corrected the total measured equivalent width,
$EW$(Li~{\sc i}+Fe~{\sc i}),
by subtracting the $EW$ of Fe~{\sc i} calculated from the empirical
relationship with ($B$--$V$) given by Soderblom et al. (1990).
The obtained values are given in  Table~\ref{tab:radvel} and plotted
in  Fig.~\ref{fig:pli} versus their spectral type.

In order to obtain an estimate of the ages of our stars
we compare their $EW$(Li~{\sc i}) with those of stars
in well-known young open clusters of different ages.  
In the $EW$(Li~{\sc i}) versus spectral type diagram (Fig.~\ref{fig:pli})
we have overplotted the upper envelope of the Li~{\sc i} $EW$ 
of IC 2602 (10-35~Myr), the Pleiades (78-125~Myr), and the Hyades (600~Myr),
open clusters which cover the range of ages of the MGs studied here.
For the Pleiades we adopt the upper envelope  determined 
by Neuh\"auser et al. (1997) with data from 
Soderblom et al. (1993b) and Garc\'{\i}a L\'{o}pez et al. (1994)
and the lower envelope given by Soderblom et al. (1993b).
In the case of IC 2602 we have not adopted the upper envelope 
given by Neuh\"auser et al. (1997) with data from 
Randich et al. (1997) and Stauffer et al. (1997) because they have used
$EW$(Li~{\sc i}) not corrected for the $EW$(Fe~{\sc i})  and we have
determined a new upper envelope with corrected $EW$(Li~{\sc i}) and 
using, in addition, new data provided by Randich et al. (2001). 
Finally for the Hyades (600~Myr) we have used the upper envelope adopted by
 Soderblom et al. (1993b).

Representative spectra in the Li~{\sc i} line region
of the star sample are plotted in Fig.~\ref{fig:li}.
As can be seen in this figure a prominent  Li~{\sc i} absorption line
is observed in the stars classified as possible members of the LA 
except GJ 211. 
The other LA stars have $EW$(Li~{\sc i}) between the 
lower and upper envelope of the Pleiades (see Fig.~\ref{fig:pli}),
except EK Dra and HD 77407 which seem to be younger
($EW$(Li~{\sc i}) between the upper envelopes
of the Pleiades and IC 2602).
The possible members of the UMa have a lower Li~{\sc i} absorption line
corresponding to the greater age of the UMa group
($EW$(Li~{\sc i})  between the  upper envelope of the Hyades and the 
lower envelope of the Pleiades).
The YD stars LQ Hya have a $EW$(Li~{\sc i}) similar to the upper envelope 
of the Pleiades.
The  $EW$(Li~{\sc i}) of GJ 577 is well above the upper envelope of the Hyades,
and no Li~{\sc i} line is detected in GJ 3706.

\begin{table*}
\caption[]{Radial velocities and Li~{\sc i} $EW$ 
\label{tab:radvel}}
\begin{flushleft}
\scriptsize
\begin{tabular}{lllrrrrrrrrr}
\noalign{\smallskip}
\hline
\noalign{\smallskip}
Name & Obs. & MJD &
$V_{\rm hel} \pm \sigma_{V}$ &
$\overline{V_{\rm hel}}$ $\pm$ $\sigma_{V}$ &
$EW$(LiI+FeI) & $EW$(LiI) & $\overline{EW{\rm (LiI)}}$ \\
\noalign{\smallskip}
     &      &     & (km s$^{-1}$) & (km s$^{-1}$)           &
(m\AA)      & (m\AA)  & (m\AA)
\scriptsize
\\
\noalign{\smallskip}
\hline
\noalign{\smallskip}
{\bf LA} \\
\noalign{\smallskip}
\hline
\noalign{\smallskip}
PW And   & 2.2m 99 & 51384.1732   & -11.76 $\pm$ 0.59 & -10.99 $\pm$ 0.11 & 290 & 269 & 271\\
         & 2.2m 99 & 51385.0401   & -11.49 $\pm$ 0.23 &                   & 283 & 262 &    \\
         & 2.2m 99 & 51386.1011   & -12.50 $\pm$ 0.37 &                   & 292 & 271 &    \\
         & 2.2m 99 & 51387.0396   & -10.51 $\pm$ 0.42 &                   & 290 & 269 &    \\
         & 2.2m 99 & 51388.1258   & -11.95 $\pm$ 0.26 &                   & 277 & 256 &    \\
         & 2.2m 99 & 51389.0818   & -11.11 $\pm$ 0.47 &                   & 296 & 275 &    \\
         & NOT 99  & 51508.8690   &  -9.41 $\pm$ 0.25 &                   & 305 & 284 &    \\
         & NOT 99  & 51509.9022   & -10.28 $\pm$ 0.28 &                   & 300 & 279 &    \\
V368 Cep & 2.2m 99 & 51384.0420   & -16.41 $\pm$ 0.39 & -16.67 $\pm$ 0.11 & 229 & 213 & 207\\
         & 2.2m 99 & 51385.0085   & -17.02 $\pm$ 0.19 &                   & 222 & 206 &    \\
         & 2.2m 99 & 51386.1725   & -16.58 $\pm$ 0.24 &                   & 231 & 215 &    \\
         & 2.2m 99 & 51387.0935   & -16.83 $\pm$ 0.32 &                   & 225 & 209 &    \\
         & 2.2m 99 & 51388.0347   & -16.74 $\pm$ 0.32 &                   & 209 & 193 &    \\
         & 2.2m 99 & 51389.0296   & -16.50 $\pm$ 0.32 &                   & 226 & 210 &    \\
         & NOT 99  & 51509.8749   & -15.72 $\pm$ 0.38 &                   & 220 & 204 &    \\
V383 Lac & 2.2m 99 & 51384.0141   & -19.55 $\pm$ 0.49 & -20.19 $\pm$ 0.12 & 266 & 249 & 257\\
         & 2.2m 99 & 51384.1565   & -19.51 $\pm$ 0.49 &                   & 265 & 248 &    \\
         & 2.2m 99 & 51384.9920   & -20.59 $\pm$ 0.22 &                   & 273 & 256 &    \\
         & 2.2m 99 & 51386.0848   & -19.98 $\pm$ 0.33 &                   & 277 & 260 &    \\
         & 2.2m 99 & 51387.0233   & -20.61 $\pm$ 0.40 &                   & 273 & 256 &    \\
         & 2.2m 99 & 51388.0023   & -20.02 $\pm$ 0.22 &                   & 277 & 260 &    \\
         & 2.2m 99 & 51388.9987   & -20.19 $\pm$ 0.36 &                   & 284 & 267 &    \\
EP Eri   & NOT 99  & 51508.9395   &  17.54 $\pm$ 0.11 &  17.54 $\pm$ 0.11 & 231 & 211 & 208\\
         & NOT 99  & 51509.9946   &                   &                   & 225 & 206 &    \\
DX Leo   & NOT 99  & 51509.2471   &   8.33 $\pm$ 0.11 &   8.13 $\pm$ 0.08 & 200 & 184 & 186\\
         & NOT 99  & 51510.2689   &   8.00 $\pm$ 0.13 &                   & 196 & 180 &    \\
         & INT 00  & 51562.1797   &                   &                   & 188 & 172 &    \\
         & INT 00  & 51564.2121   &   8.16 $\pm$ 0.24 &                   & 196 & 180 &    \\
         & INT 00  & 51566.1720   &   7.63 $\pm$ 0.24 &                   & 183 & 167 &    \\
GJ 211   & INT 00  & 51566.0980   &   0.26 $\pm$ 0.17 &   0.26 $\pm$ 0.17 & 21  & 2   & 2   \\
HD 77407 & INT 00  & 51564.1974   &   4.72 $\pm$ 0.23 &   4.43 $\pm$ 0.17 & 168 & 160 & 162\\
         & INT 00  & 51566.1601   &   4.08 $\pm$ 0.25 &                   & 173 & 165 &    \\
EK Dra   & INT 00  & 51563.3055   & -19.37 $\pm$ 0.48 & -20.64 $\pm$ 0.33 & 198 & 189 & 195\\
         & INT 00  & 51566.3100   & -21.80 $\pm$ 0.46 &                   & 201 & 202 &    \\
\noalign{\smallskip}
\hline
\noalign{\smallskip}
{\bf UMa} \\
\noalign{\smallskip}
\hline
\noalign{\smallskip}
V834 Tau & NOT 99  & 51509.0815   &   0.59 $\pm$ 0.17 &   0.27 $\pm$ 0.11 &  91 &  65 & 60 \\
         & NOT 99  & 51510.1062   &   0.07 $\pm$ 0.16 &                   &  86 &  60 &    \\
         & INT 00  & 51566.0563   &  -0.13 $\pm$ 0.35 &                   &  80 &  54 &    \\
$\pi^{1}$ UMa&NOT 99& 51510.2862  & -14.87 $\pm$ 0.15 & -14.45 $\pm$ 0.13 & 117 & 107 & 106\\
         & INT 00  &  51564.1889  & -12.98 $\pm$ 0.33 &                   & 116 & 106 &    \\
         & INT 00  &  51566.1501  & -13.82 $\pm$ 0.34 &                   & 115 & 105 &    \\
GJ 503.2 & INT 00  &  51566.2634  &  -9.26 $\pm$ 0.29 &  -9.26 $\pm$ 0.29 & 101 &  92 &  92\\
\noalign{\smallskip}
\hline
\noalign{\smallskip}
{\bf Others} \\
\noalign{\smallskip}
\hline
\noalign{\smallskip}
LQ Hya   & NOT 99  & 51509.2635   &   6.80 $\pm$ 0.41 &   8.26 $\pm$ 0.19 & 259 & 237 & 243\\
         & NOT 99  & 51510.3008   &   7.97 $\pm$ 0.29 &                   & 277 & 255 &    \\
         & INT 00  & 51564.1761   &  10.30 $\pm$ 0.50 &                   & 255 & 233 &    \\
         & INT 00  & 51566.1127   &   8.82 $\pm$ 0.37 &                   & 268 & 246 &    \\
GJ 577   & 2.2m 99 & 51384.8466   &  -6.45 $\pm$ 0.15 &  -6.48 $\pm$ 0.10 & 160 & 148 & 145\\
         & 2.2m 99 & 51386.8451   &  -6.52 $\pm$ 0.27 &                   & 165 & 153 &    \\
         & 2.2m 99 & 51388.8314   &  -6.49 $\pm$ 0.14 &                   & 145 & 133 &    \\
GJ 3706  & INT 00  & 51564.2814   &  -2.60 $\pm$ 0.21 &  -2.60 $\pm$ 0.21 & 18  & 1   &  1   \\
\noalign{\smallskip}
\hline
\end{tabular}

\end{flushleft}
\end{table*}

\section{Chromospheric activity indicators}

The echelle spectra analysed in this paper allow us to study
the behaviour of the different optical
chromospheric activity indicators
from the Ca~{\sc ii} H \& K to the Ca~{\sc ii} IRT lines,
formed at different atmospheric heights.
As shown in our previous work (Montes et el. 2000b, and references therein) 
with the simultaneous analysis of the different optical
chromospheric activity indicators and using the spectral subtraction technique,
it is possible to study in detail the chromosphere, discriminating between
the different structures: plages, prominences, flares and microflares.

\begin{figure*}
{\psfig{figure=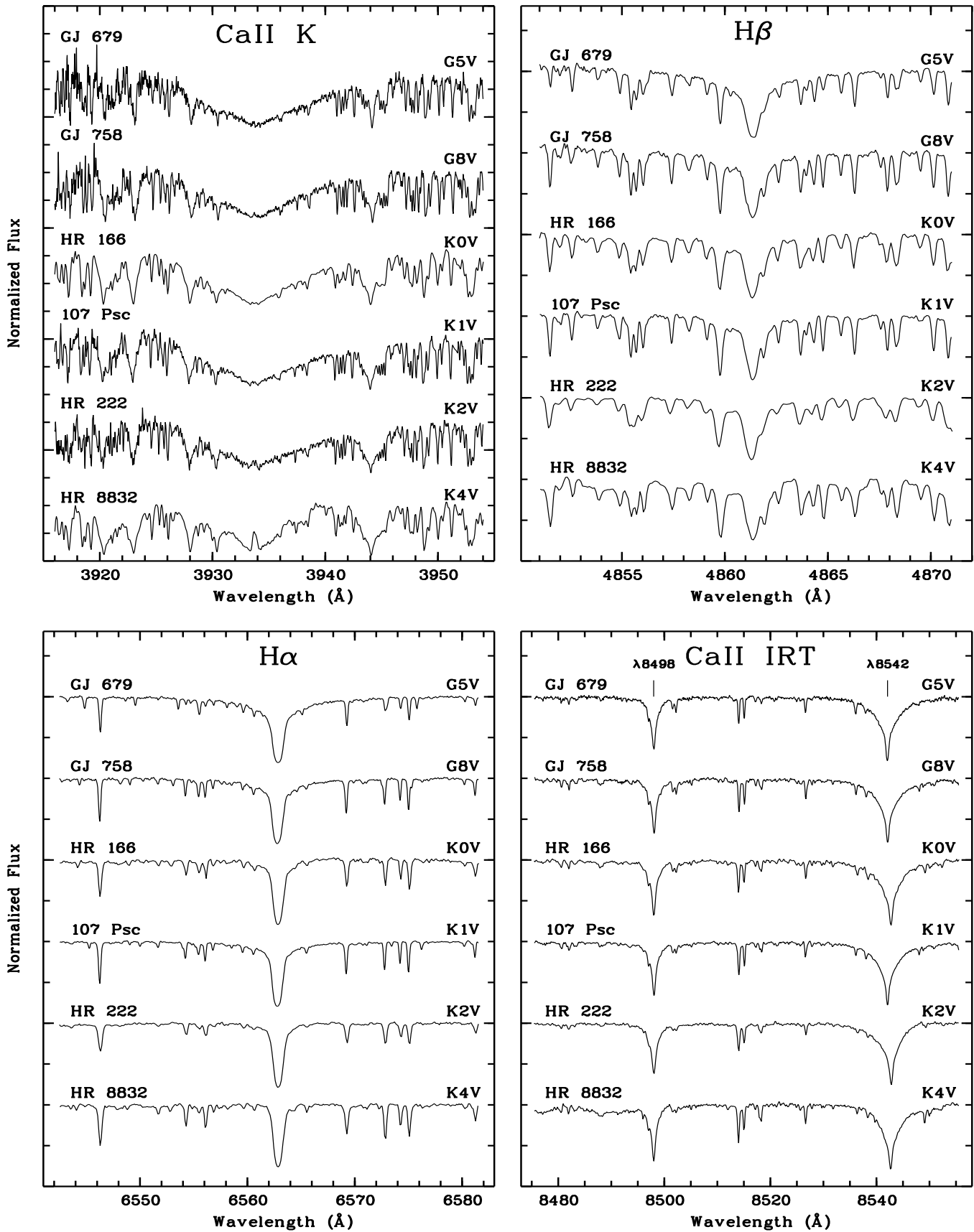,bbllx=35pt,bblly=35pt,bburx=543pt,bbury=767pt,height=22.8cm,width=18.0cm,clip=}}
\caption[ ]{Spectra of representative reference stars in the 
Ca~{\sc ii} H \& K, H$\beta$, H$\alpha$, and Ca~{\sc ii} IRT lines region.
\label{fig:ref} }
\end{figure*}

The chromospheric contribution in
these features has been determined  using
the spectral subtraction technique described in
detail by Montes et al. (1995; 1997; 1998, 2000b).
The synthesized spectrum was constructed using the program STARMOD
developed at Penn State University (Barden 1985) and modified by us. 
The inactive stars used as reference stars in the spectral subtraction
were observed during the same observing run as the active stars.
Spectra of representative reference stars in the
Ca~{\sc ii} H \& K, H$\beta$, H$\alpha$, and Ca~{\sc ii} IRT line regions
are plotted in Fig.~\ref{fig:ref}.
In Table~\ref{tab:actind}
we give the excess emission equivalent width ($EW$) (measured in the
subtracted spectra) for the Ca~{\sc ii} H \& K, H$\epsilon$,
H$\delta$, H$\gamma$, H$\beta$, H$\alpha$, and  Ca~{\sc ii} IRT
($\lambda$8498, $\lambda$8542, $\lambda$8662) lines, 
as well as the reference star used
in the subtraction technique for each observation.
We have estimated the errors in the measured $EW$ taking into account
the typical internal precisions of STARMOD
(0.5 - 2 km s$^{-1}$ in velocity shifts, and
$\pm$5 km s$^{-1}$ in  $v\sin{i}$),
the rms obtained in the fit between observed and
synthesized spectra in the regions outside the chromospheric features
(typically in the range 0.01-0.03)
 and the standard deviations resulting in the
$EW$ measurements. The estimated errors are in the range of 10-20\%.
For low active stars errors are larger and
we have considered as a clear detection of excess emission or absorption
in the chromospheric lines  only
when these features in the difference spectrum are
larger than 3$\sigma$.
Errors in the chromospheric features of the blue spectral region are 
larger due to the lower S/N of the spectra in this region. 
As an indication of the accuracy of the data, we give in
 Table~\ref{tab:obslog} the S/N in 
the Ca~{\sc ii} H \& K, and H$\alpha$ line regions.
The excess emission EW have been converted to absolute chromospheric flux at
the stellar surface by using the calibration
of Hall (1996) as a function of (B--V).
In Table~\ref{tab:actflux}
we give the absolute flux at the stellar surface
(log$F_{\rm S}$) for the lines listed in
Table~\ref{tab:actind}.

\begin{figure*}
{\psfig{figure=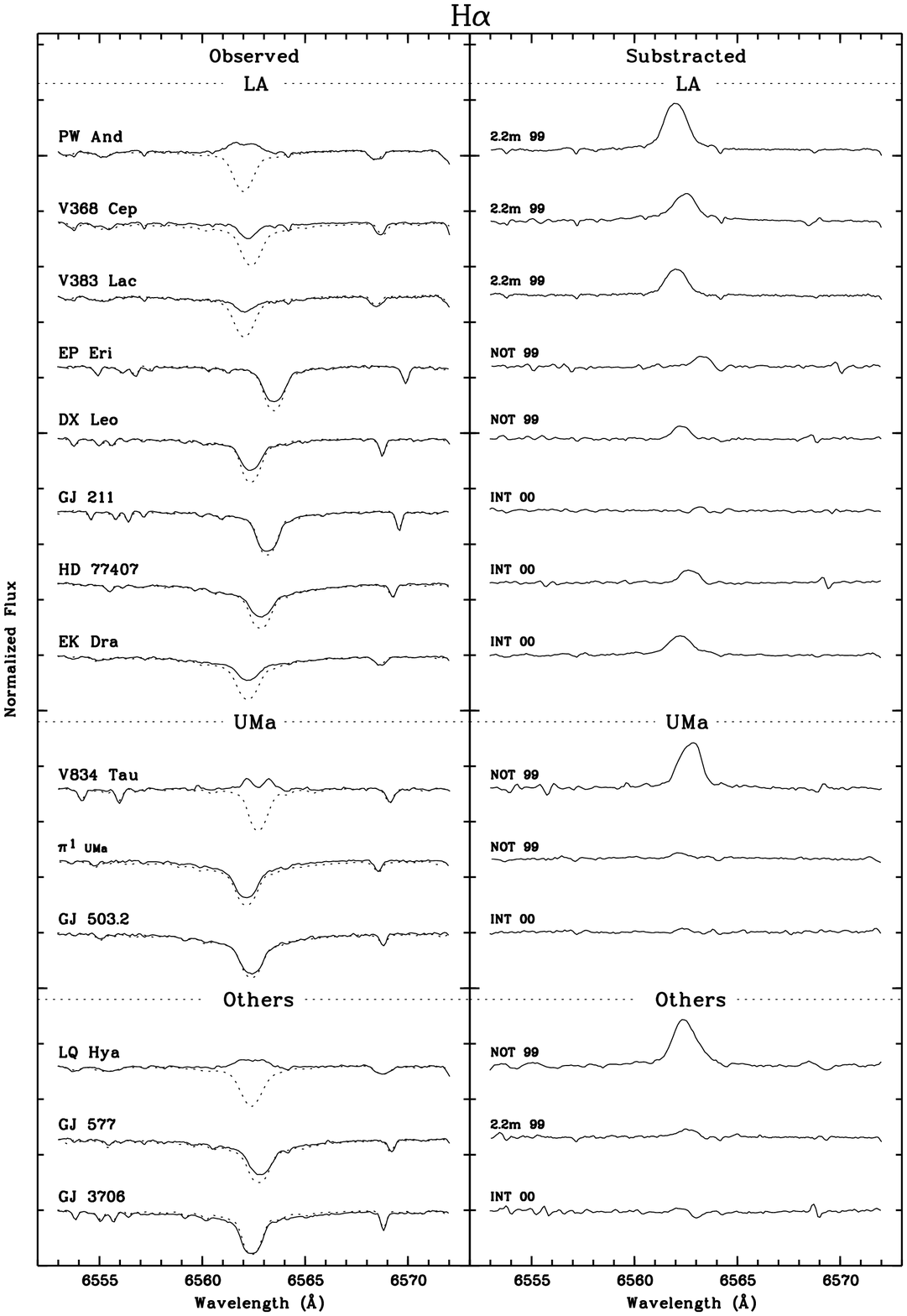,bbllx=28pt,bblly=28pt,bburx=570pt,bbury=780pt,height=22.8cm,width=18.5cm,clip=}}
\caption[ ]{Spectra in the H$\alpha$ line region for our star sample.
Observed and synthetic spectra are shown in the left panel 
and subtracted spectra in the right panel.
\label{fig:ha} }
\end{figure*}

Representative spectra in the H$\alpha$
and Ca~{\sc ii} IRT ($\lambda$8498, $\lambda$8542) line regions
of the star sample have been plotted in Figs.~\ref{fig:ha} \& ~\ref{fig:irt}.
For each star we have plotted the observed spectrum (solid-line) and the
synthesized spectrum (dashed-line) in the left panel
and the subtracted spectrum in the right panel.
H$\alpha$ emission above the continuum is detected in 
PW And, V834 Tau, and LQ Hya; 
in the rest of the stars excess H$\alpha$ emission is detected in
the subtracted spectra except GJ 3706.
Filled-in absorption in other Balmer lines is also detected in many of the
stars.
Ca~{\sc ii} H \& K emission is observed
in all the stars in which these lines are included in our spectra.
Emission reversal in the Ca~{\sc ii} IRT lines is observed in
PW And, V368 Cep, V383 Lac, DX Leo, EK Dra, V834 Tau, and LQ Hya,
in the rest of the stars a filled-in absorption line profile is observed.

\section{Individual Results}

In this section we describe the individual results about 
stellar parameters, kinematic, lithium,
chromospheric activity, and membership of the different SKG
for each star of the sample.

\subsection{PW And (HD 1405)} 

This solar neighborhood Pleiades-age K2 dwarf (Ambruster et al. 1998)
is a fast rotator with a photometric period
$P_{\rm phot}$ = 1.745 days (Hooten \& Hall 1990) 
and $v\sin{i}$ = 23.4 km s$^{-1}$ (Fekel 1997).
Strassmeier et al. (1988) listed this star as a 
 chromospherically active binary candidate, however, 
Griffin (1992) found no evidence of variability in their
radial velocities, indicating it is a single star.  
The mean radial velocity determined by us ($v_{\rm r} = -10.99$ km~s$^{-1}$)
is very similar to the values given by
Fehrenbach \& Burnage (1982) ($v_{\rm r} = -11.5$ km s$^{-1}$)
and Griffin (1992) ($v_{\rm r} = -10.4$ km s$^{-1}$)
supporting the single nature of this star.
We have found, however, small amplitude radial velocity variations 
which follow the rotational period (1.75 days) of the star.
These variations can be associated with photospheric
spots that disturb the line profile of this rapidly rotating star.
Bidelman (1985) and Christian et al. (2001) 
reported moderate Ca~{\sc ii} H \& K emission 
and the H$\alpha$ line in emission in this star. 
Chromospheric and transition region UV emission fluxes have been reported
by Ambruster et al. (1998) and Wood et al. (2000). 
In addition, it has been detected by the ROSAT-satellite as the 
2RE J001820+305 source 
(Pye et al. 1995; Kreysing et al. 1995; Thomas et al. 1998; 
Rutledge et al. 2000),
and by the EUVE-satellite as the EUVE J0018+309 source
(Malina et al. 1994; Christian et al. 2001).
In the eight spectra of this star that we have analysed, 
we have found  intense emission in the
Ca~{\sc ii} H \& K and H$\epsilon$ lines, 
strong H$\alpha$ emission above the continuum (see Fig.~\ref{fig:ha}),
excess chromospheric emissions in the other Balmer lines,
and emission reversal in the Ca~{\sc ii} IRT lines (Fig.~\ref{fig:irt}). 
We have detected variations in the excess emission of the different 
chromospheric lines, specially in the Balmer lines. 
High Li~{\sc i} abundance indicative of a Pleiades-age star 
is reported by Ambruster et al. (1998).
The $EW$(Li~{\sc i}) = 271~m\AA$\ $ we have determined confirms
it is a young star.
The space motions and all the other spectroscopic properties
we have analysed prove its LA membership.

\subsection{V368 Cep (HD 220140, HIP 115147)}

\begin{figure*}
{\psfig{figure=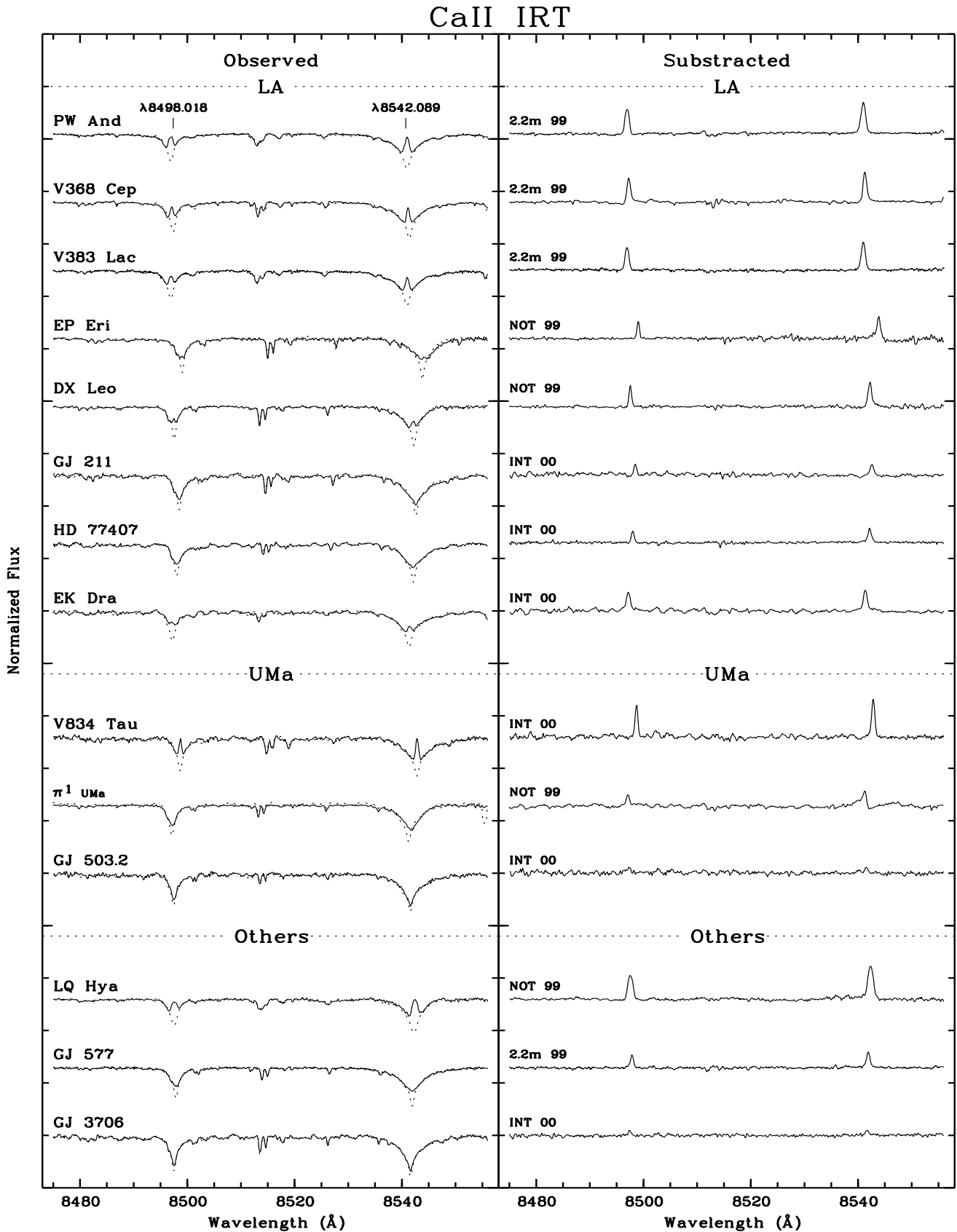,bbllx=28pt,bblly=28pt,bburx=570pt,bbury=780pt,height=22.8cm,width=18.5cm,clip=}}
\caption[ ]{Spectra in the Ca~{\sc ii} IRT (8498, 8542~\AA) line region
for our star sample.
Observed and synthetic spectra in the left panel and subtracted spectra
in the right panel.
\label{fig:irt} }
\end{figure*}

This star is the optical counterpart of the X-ray source H 2311+77 
(Pravdo et al. 1985) and
has been identified as member of the LA and classified as a
post T Tauri star (Chugainov et al. 1991a, 1993; Ambruster et al. 1998;
Kahanp\"a\"a et al. 1999). 
The spectral type of this star in the literature ranges from G5 to K2V, but
the recent photometric observations of Kahanp\"a\"a et al. (1999) 
support a K1V spectral type.
It is a rapidly rotating and spotted star with a photometric period 
$P_{\rm phot}$ = 2.74 d (Kahanp\"a\"a et al. 1999). 
and $v\sin{i}$ = 16.1 km s$^{-1}$ (Fekel 1997).
From our six spectra taken in July 1999 and another one taken in November 1999
 we have determined a constant  radial velocity
with a mean value of  -16.67 km s$^{-1}$, which is within the 
range given in the literature (15--17 km s$^{-1}$, Chugainov et al. 1991a), 
supporting its classification as a constant-velocity star.
Evidence of magnetic activity including 
strong Ca~{\sc ii} H \& K emission,
IUE-, EUVE- and ROSAT-satellite detections have been reported for this star
(Bianchi et al. 1991; Malina et al. 1994; Pye et al. 1995). 
In our spectra we observe intense emission in the
Ca~{\sc ii} H \& K and H$\epsilon$ lines, 
strong and variable excess chromospheric emissions in 
the Balmer lines, specially in H$\alpha$ (Fig.~\ref{fig:ha}), 
and the Ca~{\sc ii} IRT lines in emission superimposed 
on the corresponding absorption (Fig.~\ref{fig:irt}).
%
Chugainov et al. (1991a) report a  Li~{\sc i} $\lambda$6707.8 \AA$\ $ line 
stronger than the Ca~{\sc i} $\lambda$ 6717 \AA$\ $ line with a 
$EW$(Li~{\sc i}) = 288~m\AA.
In our spectra the Li~{\sc i} absorption feature is also stronger than 
the Ca~{\sc i} line but the mean $EW$ we have obtained after the 
correction of the Fe~{\sc i} line is $EW$(Li~{\sc i}) = 207~m\AA$\ $
which is similar to $EW$(Li~{\sc i}) observed in Pleiades stars of this
spectral type (Fig.~\ref{fig:pli}).
However, this $EW$(Li~{\sc i}) is not high enough to consider the star
as a post T Tauri star (see the $EW$(Li~{\sc i}) vs. spectral type diagram
by Mart\'{\i}n 1997).
Both kinematic and spectroscopic criteria indicate V368 Cep is
a bona fide member of the LA.

\subsection{V383 Lac (BD+48 3686)}

Recent spectroscopic and photometric studies of V383 Lac
(Mulliss \& Bopp 1994; Jeffries 1995; Henry et al. 1995; Fekel 1997; 
Osten \& Saar 1998)
concluded it is a single active K1V star with an age 
less than of the Pleiades and with a rapid rotation.
These authors report a photometric period
$P_{\rm phot}$ = 2.42 days 
and $v\sin{i}$ ranging from 14 to 20 km s$^{-1}$.
The mean radial velocity determined from our seven spectra
($v_{\rm r}$ = -20.19 km s$^{-1}$)  shows no evidence of variability
and it is in agreement with the range of values 
(from -19.4 to -22.1 km s$^{-1}$)
given in the literature, 
supporting the conclusion that it is a single star.
Mulliss \& Bopp (1994) found the H$\alpha$ and Ca~{\sc ii} IRT lines 
filled by emission.
It has been detected as an extreme ultraviolet source
(Pye et al. 1995; Lampton et al. 1997).
In our spectra, which cover more than one stellar rotation, 
we have found notable  emission in the
Ca~{\sc ii} H \& K and H$\epsilon$ lines,
excess chromospheric emission in the Balmer lines and
emission reversal in the Ca~{\sc ii} IRT lines.
During one of the nights 
a noticeable increase in the excess emission is detected,
showing the H$\alpha$ line large emission wings.
This variation could be due to a small-scale flare or to the transit of
an active region.
We have determined a $EW$(Li~{\sc i}) = 257~m\AA, similar to the values 
of 250 and 277~m\AA\ given by Mulliss \& Bopp (1994) and Jeffries (1995)
respectively. This high $EW$(Li~{\sc i}) (close to the upper envelope 
of the Pleiades) indicates it is a young star. 
The space motions and all the other spectroscopic characteristic we have
analysed in this star  confirm
it is a member of the LA.

\subsection{EP Eri (HD 17925, GJ 117, HIP 13402)}

This is a very nearby (8 pc), very young (high Li~{\sc i} abundance) active
K2-type dwarf with a rotation period of 6.85 days 
(Cutispoto 1992; Henry et al. 1995).
The presence of an unresolved companion in this star has been suggested 
by Henry et al. (1995) based on the variable widths 
of the photospheric absorption lines reported in the literature
($v\sin{i}$ range from 3 to 8; see Fekel (1997)).
Wood et al. (2000) also suggest that an unresolved secondary 
can be contributing to the emission Mg~{\sc ii} h and k lines.
However, no evidence or velocity variability is reported in the literature
(Halbwachs et al. 2000) 
and the radial velocity we have determined (17.5 km s$^{-1}$) 
is in good agreement with that of 
Beavers \& Eitter (1986) 18.8 km s$^{-1}$, 
and Henry et al. (1995) 18.1 km s$^{-1}$.
Cutispoto et al. (2001) also indicate that the binary hypothesis 
does not seem to be consistent with the Hipparcos photometric data.
Strong Ca~{\sc ii} H \& K emission and a filled-in H$\alpha$
line have been found by Pasquini et al. (1988) and Henry et al. (1995).
Chromospheric and transition region UV emission fluxes have been reported
by Ambruster et al. (1998) and Wood et al. (2000).
It is also an X-ray and EUV source (Favata et al. 1995; Jeffries 1995; 
Lampton et al. 1997).
In our spectra we have found Ca~{\sc ii} K emission and 
excess chromospheric emission in the H$\alpha$ and the Ca~{\sc ii} IRT lines.
EP Eri is a young star as indicates the strong lithium line detected  
by Cayrel de Strobel \& Cayrel (1989).
We have measured a $EW$(Li~{\sc i}) = 208~m\AA, similar 
to the 197~m\AA$\ $ given by Favata et al. (1995) and the 205~m\AA$\ $ obtained
by Jeffries (1995).
Recently, additional evidence of youth has been reported.
An age of 80 Myr has been estimated by Lachaume et al. (1999) and
a IR excess (ISO 60 $\mu$m)
has been detected in this star and attributed by Habing et al. (2001)
to a circumstellar disk (Vega-like).
Ambruster et al. (1998) identified it as member of the LA,
and the kinematic data analysed by Cayrel de Strobel \& Cayrel (1989)
show that the birth-place of this star is associated with the
Scorpio-Centaurus complex. 
The position in the (U, V) ad (U, W) planes we have determined
as well as the spectroscopic criteria 
are in agreement with its LA membership.

\subsection{DX Leo (HD 82443, GJ 354.1, HIP  46843)}

This is a nearby, young and active K0 dwarf 
with space motion very similar to the LA 
(Soderblom \& Clements 1987; Ambruster et al. 1998; Gaidos et al. 2000).
Optical flux modulation with a period of 5.4 days attributed to 
cool photospheric spots have been found by
Henry et al. (1995), Messina et al. (1999b), and  Gaidos et al. (2000).
A projected rotational velocity $v\sin{i}$ = 6.2 is given by
Fekel (1997).
DX Leo is a single star as indicates the constant radial velocity
we have determined in our spectra, $v_{\rm r}$ = 8.13 km s$^{-1}$, and
the values of $v_{\rm r}$ = 8.2, 8.9, and 8.25 km s$^{-1}$ given by 
Duquennoy et al. (1991), Griffin (1994), and Henry et al. (1995)
respectively. 
Strong chromospheric and transition region line emissions
have been reported by Soderblom \& Clements (1987), 
Basri et al. (1989), Ambruster et al. 1998; 
Strassmeier et al. (2000), and Wood et al. (2000).
It is also a X-ray and EUV source 
(Pye et al. 1995; H{\" u}nsch et al. 1999).
In our spectra, taken in two different epochs, we have found 
noticeable emission in the Ca~{\sc ii} H \& K and  Ca~{\sc ii} IRT lines,
and excess chromospheric emission in the H$\alpha$ line.
We have obtained a $EW$(Li~{\sc i}) = 198~m\AA$\ $ 
which is within the range observed in the Pleiades and similar to
the value of $EW$(Li~{\sc i}) = 187~m\AA$\ $ 
given by Strassmeier et al. (2000).
The space motion we have determined and all the spectroscopic criteria
we have analysed indicate DX Leo is a young star member of the LA. 

\subsection{GJ 211 (HD 37394, HR 1925)}

This is a nearby star classified as a possible member of the LA
(Jeffries \& Jewell 1993; Gaidos et al. 2000).
It is a K1V slowly rotating star ($P_{\rm phot}$ = 10.86 d, 
and $v\sin{i}$ = 4.0 km s$^{-1}$, Gaidos et al. 2000) 
but with evidence of chromospheric activity 
(emission in the Ca~{\sc ii} H \& K and Mg~{\sc ii} h \& k lines, 
Soderblom \& Clements 1987) and coronal activity 
(EUV and X-ray emission  H{\" u}nsch et al. 1999).
It has even been classified by some authors (Gershberg et al. 1999) 
as a flare star.
Unlike EP Eri, Habing et al. (2001) have not detected
evidence of circumstellar disk in the IR (ISO 60 $\mu$m)
flux of  this star.
Constant radial velocity has been reported for this star with values 
ranging from -0.2 to 2 km s$^{-1}$, which are in agreement with 
the $v_{\rm r} = 0.26 $ km s$^{-1}$ we have determined.
In our spectrum we observe small excess  chromospheric emission in
the Balmer lines and Ca~{\sc ii} IRT lines
and a very small $EW$(Li~{\sc i}) (2.0 m\AA) in agreement with 
the $EW$(Li~{\sc i}) = 1.3$\pm$3.2 m\AA$\ $ reported by Gaidos et al. (2000).
The space motions calculated by us are consistent with the LA but
the low level of chromospheric emission and very small $EW$(Li~{\sc i})
indicate it is not a young star and probably it is not a member of the LA. 

\subsection{HD 77407 (BD+38 1993, HIP~44458)}

This G0 star was included in the study  
of the Hyades and Sirius MGs by Eggen (1986),
but not identified as a member of any of these MGs. 
It is an X-ray/EUV source detected by ROSAT and EUVE
(Lampton et al. 1997)  
and also detected as a stellar radio source by Helfand et al. (1999).
However, it is a very little-studied star, and no previous determinations of
radial velocity, rotation, chromospheric activity and lithium have been 
reported in the literature.
Using our two spectra of this star we have determined a 
mean radial velocity of 4.43 km s$^{-1}$, which together with the
astrometric data results in a Galactic space motion ($U$, $V$, $W$) 
similar to the LA. Eggen's kinematic criteria also confirm
their membership of the LA.
Our spectra show that it is a slow rotating star 
($v\sin{i}$ $\approx$ 7 km s$^{-1}$)
with a notable chromospheric excess emission in the $H\beta$, $H\alpha$, 
and Ca~{\sc ii} IRT lines. 
The $EW$(Li~{\sc i}) = 162~m\AA$\ $ we have determined lies above the upper
$EW$(Li~{\sc i}) envelope of the Pleiades 
(as also reported by Wichmann \& Schmitt 2001), 
indicating that it is a very young star and 
therefore a probable member of the LA. 

\subsection{EK Dra (HD 129333, GJ 559.1A, HIP  71631)}

This nearby G1.5V spotted and very active star
(Strassmeier \& Rice 1998; Strassmeier et al. 2000) 
was previously identified as a member of the LA  
(Chugainov 1991; Chugainov et al. 1991b; Soderblom \& Clements 1987) 
and considered as a young solar analog (Dorren \& Guinan 1994). 
It is a rapidly-rotating star with $v\sin{i}$ 17.3 km s$^{-1}$ and
with a photometric period ranging in the literature
from 2.6 to 2.8 days (Chugainov et al. 1991b; Dorren \& Guinan 1994; 
Strassmeier \& Rice 1998; Messina et al. 2001; DePasquale et al. 2001).
EK Dra can be treated essentially as a single star, although 
Duquennoy \& Mayor (1991) have suggested that it may be a member 
of a long-period ($\approx$ 12.5~yr) binary system 
with radial velocity variations 
between -21 and -32 km s$^{-1}$ and $v_{\rm 0}$ = -23.1 km s$^{-1}$.
Dorren \& Guinan (1994) found a mean radial velocity of -24 km s$^{-1}$,
Fehrenbach et al. (1997) give -18 km s$^{-1}$ and
we have obtained in our spectra a mean value of -20.6 km s$^{-1}$.
High levels of magnetic activity have been detected in this star:
strong Ca~{\sc ii} H \& K emission (Soderblom 1985);
variable UV chromospheric emission lines 
(Dorren \& Guinan 1994; Saar \& Bookbinder 1998); and
X-ray and EUV emission (G{\" u}del et al. 1997;  Audard et al. 2000).
In our spectra we observe a small emission reversal 
in the Ca~{\sc ii} IRT lines and
a notable excess emission in the $H\alpha$ and $H\beta$ lines.
Strong Li~{\sc i} absorption has been reported previously in
the literature for this star (see Wichmann \& Schmitt 2001).
In our spectra we have determined a 
$EW$(Li~{\sc i}) of 198~m\AA, which is
 between the upper envelopes of the Pleiades and IC 2602,
indicating that this star is significantly younger than the Pleiades
open cluster.
The space motion, high level of magnetic activity, and strong Li~{\sc i}
absorption observed in this star confirm that it is a member of the LA. 

\subsection{V834 Tau (HD 29697, GJ 174, HIP  21818)}

This very active K4V star was
considered for some time (Strassmeier et al. 1993)
as candidate to chromospherically active binary.
However, the constant radial velocities measured by several authors
(Fouts \& Sandage 1986; Henry et al. 1995; Halbwachs et al. 2000) 
and our radial velocity determination ($v_{r} = 0.27$  km s$^{-1}$) 
indicate that it is a single star. 
Light variations were first discovered in this star by Chugainov (1981),
later, Henry et al. (1995) determined 
a photometric period of 3.936 days and Fekel (1997) measured 
a rotational velocity, $v\sin{i}$ = 9.5 km s$^{-1}$.
Exceptional strong Ca~{\sc ii} H \& K emission lines (Young et al. 1989),  
modest H$\alpha$ emission above the continuum 
(Rutten et al. 1989; Henry et al. 1995) and
Radio, X-ray and EUV emission 
(G{\" u}del 1992; Pye et al. 1995; H{\" u}nsch et al. 1999)
have been detected in this star.
In our spectra taken at two different epochs we have found 
strong emission in the Ca~{\sc ii} K line, the H$\alpha$ line in 
emission above the continuum with a central self-absorption, 
a filled-in absorption $H\beta$ line, 
and the Ca~{\sc ii} IRT lines in emission.
Chugainov (1991) and Eggen (1996) listed V834 Tau as a young disk star 
and Chugainov (1991) suggested it as a possible member of the LA.
However, the Galactic velocity components we have determined
indicate it is a possible member of the UMa.
The lithium line has been detected in this star; Henry et al. (1995) 
give a $EW$(Li~{\sc i}) = 79~m\AA, and we have determined
in our spectra a $EW$(Li~{\sc i}) = 60~m\AA.
This notable $EW$(Li~{\sc i}) (see Fig.~\ref{fig:pli}) indicates 
it is a young star and confirms its membership of the UMa.

\subsection{$\pi^{1}$ UMa (HD 72905, GJ 311, HIP  42438)}

This nearby G1.5V active star, considered as proxy of the young Sun
(Bochanski et al. 2001), has been
classified as a possible member of the UMa moving group by
Soderblom \& Clements (1987); Soderblom \& Mayor (1993a); 
Gaidos (1998); and Gaidos et al. (2000). 
It has a $v\sin{i}$ = 9.5 km s$^{-1}$ (Fekel 1997) 
and a short rotation period (P $\approx$ 4.8 days) as measured
by periodic light variations due to starspots by  
Gaidos et al. (2000) and Bochanski et al. (2001). 
The mean radial velocity we have determined 
($v_{\rm r}$ = -14.45 km s$^{-1}$) is similar to the value 
reported by Duquennoy et al. (1991) ($v_{\rm r}$ = -12.66 km s$^{-1}$).
$\pi^{1}$ UMa has 
high levels of chromospheric and coronal activity 
(Soderblom \& Mayor 1993a, b; Dorren \& Guinan 1994).
A superflare (Schaefer et al. 2000) was detected in this star 
in the X-ray band by the EXOSAT satellite (Landini et al. 1986).
In our observations we have found small excess chromospheric emissions in
the H$\alpha$, and the Ca~{\sc ii} IRT lines.
We have determined a $EW$(Li~{\sc i}) = 106~m\AA,
very close to the value of 96~m\AA\ given by Soderblom et al. (1993a).
This $EW$(Li~{\sc i}) is intermediate between 
the Hyades upper envelope and Pleiades lower envelope, corresponding to
the age of the UMa.
Our new calculation of the galactic velocity components and
the spectroscopic criteria above-described 
are in agreement with the membership of this star to the UMa.

\subsection{GJ 503.2 (HD 115043, BD+57 1425, HIP  64532)}

This nearby G2V star is classified as a member of the
UMa group by Eggen (1992) and Soderblom \& Mayor (1993a).
It is a slow rotating star with $v\sin{i}$ = 7.5  km s$^{-1}$ 
(Soderblom \& Mayor 1993b).
%
%
Duquennoy et al. (1991) reported a constant radial velocity 
($v_{\rm r}$ = -8.86 km s$^{-1}$) for this star, which is very similar to
the value determined in our spectrum ($v_{\rm r}$ = -9.26 km s$^{-1}$)
confirming it is a single star.
Evidence of magnetic activity has been found in the X-ray 
by the ROSAT (H{\" u}nsch et al. 1999) and in the
ultraviolet by the IUE (Soderblom \& Clement 1987) and the HST (Lamzin 2000).
Moderate Ca~{\sc ii} H \& K chromospheric emission is reported by
Soderblom (1985).
In our spectra we have found a very small fill-in 
in the H$\alpha$ and Ca~{\sc ii} IRT lines.
The Galactic velocity components we have determined and the 
kinematic criteria are compatible with the star being a member of the UMa.
In addition, the $EW$(Li~{\sc i}) of 92~m\AA$\ $ determined in our
spectrum, which is similar to the 
value of 77~m\AA$\ $ given by Soderblom et al. (1993a), 
indicates an age intermediate between the Hyades and Pleiades 
corresponding to the UMa.
All this supports that GJ 503.2 is
a bona fide member of the UMa.

\subsection{LQ Hya (HD 82558, GJ 355, HIP  46816)}

This star is a young single K2 dwarf classified as 
a BY Dra variable (Fekel et al.~1986).
It is a rapidly rotating star with 
{\it v} sin{\it i} = 26.5 km s$^{-1}$ (Donati 1999)
and with a photometric rotational period 
of 1.600881 days (Strassmeier et al. 1997).
It is a very active star, as indicated by emission
in several chromospheric and transition region lines, 
even with occasional flares 
(see Montes et al. 1999 and references therein).
EUV and X-ray emission and X-ray flares have been detected in this 
star (see Covino et al. 2001 and references therein).
In our four spectra taken at two different epochs we observed
strong emission in the Ca~{\sc ii} K line, the H$\alpha$ line in
emission above the continuum with an intensity similar to that observed 
in the quiescent spectra by Montes et al. (1999),
a filled-in absorption $H\beta$ line,
and the Ca~{\sc ii} IRT lines in emission.
We have determined a mean radial velocity, $v_{\rm r}$ = 8.26 km s$^{-1}$, 
very close to the mean value of 
7.3 km s$^{-1}$ reported by Fekel et al. (1986), 
7.5 km s$^{-1}$ given by Vilhu et al.~(1991), and
9.0  km s$^{-1}$ by Donati et al. (1997),
confirming it is a constant radial velocity star. 
Eggen (1984b) suggested that LQ Hya may be a member of the HS,
Fekel et al. (1986), using their new radial velocity, found it should be 
considered only as a YD star, however,
Chugainov (1991) and  Ambruster et al. (1998) 
listed this star as a member of the LA. 
The U, V and W velocity components we have calculated 
using the astrometric data from Tycho-2 Catalogue and the radial velocity
determined by us
indicate that this star is a YD star
but not a member of the LA. 
It is, however, a young star as pointed out by the strong lithium absorption 
line ($EW$(Li~{\sc i} = 234 ~m\AA) reported by Fekel et al. (1986).
In our spectra we have obtained a similar mean  
$EW$(Li~{\sc i} of 243~m\AA, which is close to the upper envelope
of the Pleiades cluster.

\subsection{GJ 577 (IU Dra, HD 134319,  HIP 73869)}

Messina \& Guinan (1998) and Messina et al. (1999a)
considered this G5V star as a proxy for the young Sun
and classified it as a probable member of the HS 
according with its $U$, $V$, $W$ components and parallax.
These authors found high levels of photospheric magnetic activity in 
this star and reported a photometric rotation period
of 4.448 days.
The mean radial velocity we have determined with our three spectra 
($v_{\rm r}$ = -6.48 km s$^{-1}$) is very close to the
constant radial velocity ($v_{\rm r}$ = -6.38 km s$^{-1}$) given
by Duquennoy et al. (1991), confirming it is likely a single star.
Soderblom (1985) found moderate chromospheric emission in this star.
Moderate Ca~{\sc ii} H \& K emission is observed in our spectra,
however, the H$\epsilon$ line is not in emission.
In addition, a notable filling-in is detected in the H$\alpha$ 
and Ca~{\sc ii} IRT lines.
The behaviour of the lithium ($\lambda$6707.8 line) in this star 
has not been previously reported in the literature.
We have determined in our spectra a $EW$(Li~{\sc i}) of 145~m\AA$\ $
which is well above the upper envelope of the Hyades and 
close to the lower envelope of the Pleiades (see Fig~\ref{fig:pli}).
Even though this star could be considered as member of the HS 
based on its position in the (U, V) plane and the kinematic criteria,
the $EW$(Li~{\sc i}) indicates it is too young 
to be a member of the HS. 

\subsection{GJ 3706 (HD 105631, BD+41 2276, HIP  59280)}

This K0V star was classified as a member of the IC 2391
supercluster by Eggen (1991).
The radial velocity we have determined for this star 
($v_{\rm r}$ = -2.6 km s$^{-1}$) is 
in agreement with the previous value ($v_{\rm r}$ = -3.1 km s$^{-1}$)
reported by Duflot et al. (1995).
This star is listed by H{\" u}nsch et al. (1999) as 
source of X-ray detected by the ROSAT-satellite and
Strassmeier et al. (2000) found slight Ca~{\sc ii} H\&K emission.
In our spectrum we have found a very low level of chromospheric activity 
(no filled-in is detected in H$\alpha$ and a very slight filled-in 
is observed in the Ca~{\sc ii} IRT lines).
The Li~{\sc i} absorption line is practically not detected
in our spectrum ($EW$(Li~{\sc i})= 1.4 m\AA)
indicating it is not a young star 
and probably not a member of the IC 2391 SC,
in spite of the fact that the kinematic criteria 
point out that it is a member of this SC.

\section{Discussion and conclusions}

In this paper we have used high resolution echelle spectroscopic observations 
to test the membership of 14 single late-type stars to young stellar 
kinematic groups such as 
the Local Association (20 - 150 Myr),
Ursa Major group (300 Myr),
 Hyades supercluster (600 Myr), and
 IC 2391 supercluster (35 Myr).
We have determined accurate heliocentric radial velocities, 
equivalent width of the Li~{\sc i} doublet at $\lambda$6707.8 \AA,
and the level of chromospheric activity using different indicators
from the Ca~{\sc ii} H \& K to the Ca~{\sc ii} IRT lines.
All these data allow us to apply  both kinematic 
(position in the (U, V) and (W, V) planes and Eggen's criteria, see Paper~I)
and spectroscopic (chromospheric activity, and $EW$(Li~{\sc i})) criteria.

Using the kinematic criteria we have classified 
PW And, V368 Cep, V383 Lac, EP Eri, DX Leo, GJ 211, HD 77407, and EK Dra
as possible members of the LA. 
The $EW$(Li~{\sc i}) and level of chromospheric activity of all 
these stars, except GJ 211, indicate ages similar to the Pleiades
or even younger than the Pleiades (HD 77407 and EK Dra)
confirming their membership to the LA.
However, the low level of activity and the Li~{\sc i} line close to
the limit of detection we have found in GJ 211 indicate an age older
than the range of ages assigned to this SKG and that this star 
should be rejected as a possible member.
 
V834 Tau, $\pi^{1}$ UMa, and GJ 503.2 turn out to be 
possible members of the UMa according to the kinematic criteria.
The spectroscopic criteria also confirm their membership.
We have found for these three stars
a moderate level of chromospheric activity, and
a $EW$(Li~{\sc i}) between the 
upper envelope of the Hyades and the lower envelope of the Pleiades,
that corresponds to the age of 300~Myr of the UMa group, which is 
intermediate between the Pleiades and Hyades.

The previously identified member of the LA, LQ Hya, 
turns out to be a YD star (i.e. space-velocity components inside the
boundaries that determine the young disk population)
but with no clear membership of any of the young SKG studied here.
However, the spectroscopic criteria confirm it is a young star
with $EW$(Li~{\sc i}) similar to the upper envelope
of the Pleiades and the H$\alpha$ line in emission above the continuum. 

GJ 577 was previously classified as a member of the HS, 
and the kinematic criteria we have applied confirm this classification.
However, the $EW$(Li~{\sc i}) well above the upper envelope of the Hyades 
that we have determined for this star indicates that it is younger than the 
age assigned to the HS.

Even though GJ 3706 could be considered as a member of IC 2391 
according to the kinematic criteria, the very small $EW$(Li~{\sc i})
we have determined in our spectrum indicates an age too old to 
be a member of the very young IC 2391 SKG.

An additional age estimation of these stars 
and those for the possible late-type stars
members of young SKG we have selected in Paper~I can be obtained
by isochrone fitting on the color-magnitude diagram.
We are carrying out this kind of study 
in our ongoing project
dedicated to the  detailed study of each young SKG
and the results will be addressed in forthcoming papers.

Some stars have been observed at different nights and at
different epochs, covering several rotational periods.
The radial velocities we have determined in these spectra 
show no evidence of variability and are in good 
agreement with the range of values previously reported by other authors,
supporting the single nature of these stars.
For some of these stars (with several spectra available) we have also found
low level variability of the chromospheric emission, which can be attributed
to low level flaring (V383 Lac) or the rotational modulation of 
chromospheric active regions (PW And, V368 Cep).

\begin{figure}
{\psfig{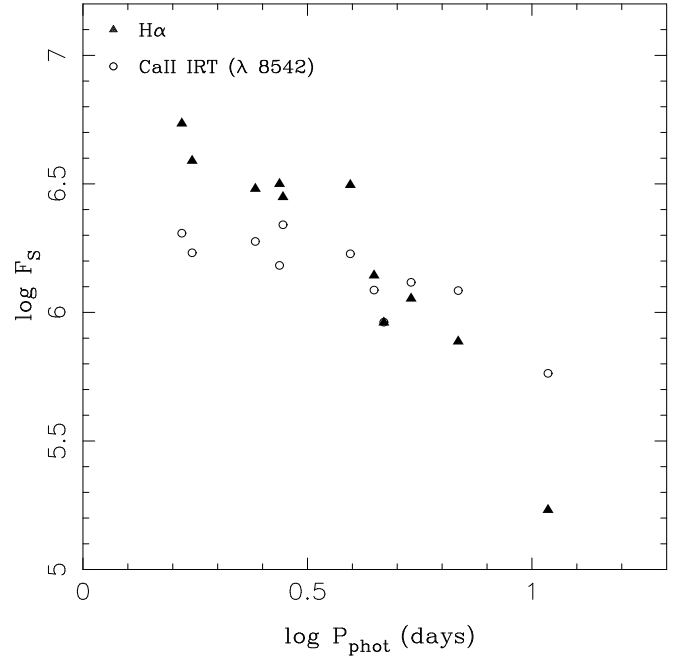}}
\caption[ ]{Absolute flux at the stellar surface
in the H$\alpha$, and Ca~{\sc ii} IRT lines
versus photometric period.
\label{fig:flux_period} }
\end{figure}

The stars with the highest levels of chromospheric activity
(LQ Hya, V834 Tau, PW And) have the
H$\alpha$ line in emission above the continuum 
and also have the highest excess emissions in the Ca~{\sc ii} H \& K 
and Ca~{\sc ii} IRT lines.
These three stars are also the most rapidly-rotating stars of the sample
with rotation period, $P_{\rm phot}$ $<$ 2 days.
When we analyse in detail the behaviour of the chromospheric excess emissions
with the star rotation 
(characterized by their photometric period, $P_{\rm phot}$ or their
projected rotational velocity, $v\sin{i}$, given in Table~\ref{tab:par})
a clear trend of increasing activity with increasing rotation is revealed.
This can be seen in Fig.~\ref{fig:flux_period}, where we have plotted 
the absolute flux at the stellar surface (log$F_{\rm S}$)
in the H$\alpha$, and Ca~{\sc ii} IRT lines
versus the photometric period (log$P_{\rm phot}$).

This behaviour confirms that this group of young stars 
also follows a rotation-activity relation similar to that observed 
in other kinds of active stars (see Montes et al. 1995), and in stars
members of young open clusters (see Simon 2001, and references therein).
Chromospheric and transition region (using UV emission lines) 
rotation-activity relations have been previously reported by 
Ambruster et al. (1998) for five of the stars of our sample.
 
A more detailed analysis 
of the relative behaviour of the different
diagnostics between themselves and with respect to the main stellar
parameters of some of these
stars including additional spectroscopic observations 
will be addressed in forthcoming papers.

We have already started a program of high resolution
echelle spectroscopic observations
of a large sample of late-type stars 
(selected by us in Paper~I as possible members of young SKGs)
in order to carry out a spectroscopic analysis similar 
to that described in this paper, and in this way
better establish their  membership of different SKGs
(for preliminary results of this spectroscopic survey see 
Montes et al. 2001c).

\begin{acknowledgements}

We would like to thank Dr. B.H. Foing for allow us to use the
ESA-MUSICOS spectrograph at Isaac Newton Telescope.
This research has made use the
of the SIMBAD data base, operated at CDS,
Strasbourg, France, and the ARI Database for Nearby Stars,
Astronomisches Rechen-Institut, Heidelberg.
We would also like to thank the referee S. Catalano
for suggesting several improvements and clarifications.
This work was supported by the Universidad Complutense de Madrid and
the Spanish Direcci\'{o}n General de Ense\~{n}anza Superior e
Investigaci\'{o}n Cient\'{\i}fica (DGESIC) under grant PB97-0259.

\end{acknowledgements}

\begin{table*}
\caption[]{EW of the different chromospheric activity indicators
\label{tab:actind}}
\begin{flushleft}
\scriptsize
\begin{tabular}{llcccccccccccc}
\noalign{\smallskip}
\hline
\noalign{\smallskip}
     &      &     & \multicolumn{10}{c}{$EW$(\AA) in the subtracted spectrum} \\
\cline{4-13}
\noalign{\smallskip}
Name & Obs. & MJD & \multicolumn{2}{c}{Ca II} & & & & & &
\multicolumn{3}{c}{Ca II IRT} & Ref. Star \\
\cline{4-5}\cline{11-13}
\noalign{\smallskip}
     &      &     &
 K   & H  & H$\epsilon$ & H$\delta$ & H$\gamma$ & H$\beta$ & H$\alpha$ &
$\lambda$8498 & $\lambda$8542 & $\lambda$8662
\scriptsize
\\
\noalign{\smallskip}
\hline
\noalign{\smallskip}
{\bf LA} \\
\noalign{\smallskip}
\hline
\noalign{\smallskip}
PW And   & 2.2m 99 & 51384.1732 &
1.78 & 1.17 & 0.31 & 0.27 & 0.29 & 0.47 & 1.39 & 0.49 & 0.63 & 0.52 & GJ 706\\
         & 2.2m 99 & 51385.0401 &
1.36 & 0.99 & 0.33 & 0.28 & 0.28 & 0.45 & 1.26 & 0.51 & 0.63 & 0.56 &   "   \\
         & 2.2m 99 & 51386.1011 &
1.98 & 0.93 & 0.32 & 0.26 & 0.29 & 0.51 & 1.35 & 0.50 & 0.66 & 0.53 &   "   \\
         & 2.2m 99 & 51387.0396 &
  -  &   -  &   -  & 0.28 & 0.27 & 0.44 & 1.23 & 0.47 & 0.62 & 0.55 &   "   \\
         & 2.2m 99 & 51388.1258 &
1.97 & 0.82 & 0.26 & 0.24 & 0.26 & 0.48 & 1.20 & 0.47 & 0.61 & 0.52 &   "   \\
         & 2.2m 99 & 51389.0818 &
2.18 & 1.68 & 0.33 &   -  &   -  & 0.47 & 1.36 & 0.49 & 0.68 & 0.53 &   "   \\
         & NOT 99  & 51508.8690 &
2.43 &   -  &   -  &   -  &   -  & 0.71 & 1.58 & 0.58 & 0.78 &   -  & HR 222\\
         & NOT 99  & 51509.9022 &
2.36 &   -  &   -  &   -  &   -  & 0.71 & 1.54 & 0.53 & 0.91 &   -  &   "   \\
V368 Cep & 2.2m 99 & 51384.0420 &
1.25 & 0.68 & 0.20 & 0.17 & 0.22 & 0.28 & 0.82 & 0.36 & 0.48 & 0.40 & GJ 758\\
         & 2.2m 99 & 51385.0085 &
  -  & 0.65 & 0.25 & 0.14 & 0.16 & 0.27 & 0.77 & 0.36 & 0.48 & 0.40 &   "   \\
         & 2.2m 99 & 51386.1725 &
0.87 & 0.59 & 0.20 & 0.17 & 0.14 & 0.23 & 0.85 & 0.34 & 0.48 & 0.40 &   "   \\
         & 2.2m 99 & 51387.0935 &
0.97 & 0.61 & 0.19 & 0.20 & 0.18 & 0.27 & 1.00 & 0.37 & 0.49 & 0.42 &   "   \\
         & 2.2m 99 & 51388.0347 &
0.93 & 0.65 & 0.23 & 0.08 & 0.08 & 0.28 & 0.64 & 0.32 & 0.43 & 0.39 &   "   \\
         & 2.2m 99 & 51389.0296 &
1.23 & 0.78 & 0.29 &   -  &   -  & 0.46 & 1.11 & 0.39 & 0.52 & 0.52 &   "   \\
         & NOT 99  & 51509.8749 &
1.07 &   -  &   -  &   -  &   -  & 0.22 & 0.55 & 0.31 & 0.65 &   -  & HR 166\\
V383 Lac & 2.2m 99 & 51384.0141 &
1.36 & 0.71 & 0.21 & 0.06 & 0.09 & 0.21 & 0.59 & 0.40 & 0.52 & 0.43 & GJ 706\\
         & 2.2m 99 & 51384.1565 &
1.24 & 0.74 & 0.23 & 0.07 & 0.07 & 0.18 & 0.61 & 0.39 & 0.53 & 0.44 &   "   \\
         & 2.2m 99 & 51384.9920 &
  -  &   -  &   -  & 0.0  & 0.04 & 0.18 & 0.64 & 0.42 & 0.59 & 0.48 &   "   \\
         & 2.2m 99 & 51386.0848 &
1.17 & 1.03 & 0.38 & 0.17 & 0.20 & 0.40 & 1.08 & 0.51 & 0.73 & 0.64 &   "   \\
         & 2.2m 99 & 51387.0233 &
1.17 & 0.74 & 0.26 & 0.08 & 0.08 & 0.30 & 0.69 & 0.40 & 0.55 & 0.46 &   "   \\
         & 2.2m 99 & 51388.0023 &
1.18 & 0.78 & 0.20 & 0.06 & 0.08 & 0.19 & 0.62 & 0.43 & 0.58 & 0.48 &   "   \\
         & 2.2m 99 & 51388.9987 &
1.31 & 0.70 & 0.19 &   -  &   -  & 0.22 & 0.72 & 0.39 & 0.57 & 0.49 &   "   \\
EP Eri   & NOT 99  & 51508.9395 &
0.67 &   -  &   -  &   -  &   -  & 0.0  & 0.19 & 0.18 & 0.39 &   -  & HR 166\\
         & NOT 99  & 51509.9946 & \\
DX Leo   & NOT 99  & 51509.2471 &
0.59 &   -  &   -  &   -  &   -  & 0.0  & 0.24 & 0.23 & 0.39 &   -  & HR 166\\
         & NOT 99  & 51510.2689 &
0.76 &   -  &   -  &   -  &   -  & 0.0  & 0.23 & 0.24 & 0.37 &   -  &   "   \\
         & INT 00  & 51562.1797 &
  -  &   -  &   -  &   -  &   -  & 0.0  & 0.22 & 0.30 & 0.35 & 0.32 &   "   \\
         & INT 00  & 51564.2121 &
  -  &   -  &   -  &   -  &   -  & 0.0  & 0.26 & 0.25 & 0.36 & 0.33 &   "   \\
         & INT 00  & 51566.1720 &
  -  &   -  &   -  &   -  &   -  & 0.0  & 0.20 & 0.24 & 0.38 & 0.28 &   "   \\
GJ 211   & INT 00  & 51566.0980 &
  -  &   -  &   -  &   -  &   -  & 0.0  & 0.04 & 0.12 & 0.18 & 0.12 & HR 166\\
HD 77407 & INT 00  & 51564.1974 &
  -  &   -  &   -  &   -  &   -  & 0.07 & 0.25 & 0.18 & 0.27 & 0.26 & Sun\\
         & INT 00  & 51566.1601 &
  -  &   -  &   -  &   -  &   -  & 0.03 & 0.19 & 0.20 & 0.23 & 0.23 &   " \\
EK Dra   & INT 00  & 51563.3055 &
  -  &   -  &   -  &   -  &   -  & 0.23 & 0.58 & 0.39 & 0.45 & 0.46 & Sun\\
         & INT 00  & 51566.3100 &
  -  &   -  &   -  &   -  &   -  & 0.22 & 0.63 & 0.36 & 0.47 & 0.44 &   " \\
\noalign{\smallskip}
\hline
\noalign{\smallskip}
{\bf UMa} \\
\noalign{\smallskip}
\hline
\noalign{\smallskip}
V834 Tau & NOT 99  & 51509.0815 &
3.62 &   -  &   -  &   -  &   -  & 0.55 & 1.51 & 0.56 & 0.91 &   -  & HR 166\\
         & NOT 99  & 51510.1062 &
3.64 &   -  &   -  &   -  &   -  & 0.50 & 1.29 & 0.74 & 0.95 &   -  &   "   \\
         & INT 00  & 51566.0563 &
  -  &   -  &   -  &   -  &   -  & 0.55 & 1.35 & 0.40 & 0.55 & 0.44 &   "   \\
$\pi^{1}$ UMa&NOT 99 & 51510.2862 &
  -  &   -  &   -  &   -  &   -  & 0.0  & 0.19 & 0.15 & 0.30 &   -  & Sun\\
         & INT 00  & 51564.1889 &
  -  &   -  &   -  &   -  &   -  & 0.0  & 0.10 & 0.13 & 0.16 & 0.19 &   " \\
         & INT 00  & 51566.1501 &
  -  &   -  &   -  &   -  &   -  & 0.0  & 0.10 & 0.13 & 0.17 & 0.18 &  "   \\
GJ 503.2 & INT 00  & 51566.2634 &
  -  &   -  &   -  &   -  &   -  & 0.0  & 0.06 & 0.09 & 0.11 & 0.12 & Sun \\
\noalign{\smallskip}
\hline
\noalign{\smallskip}
{\bf Others} \\
\noalign{\smallskip}
\hline
\noalign{\smallskip}
LQ Hya   & NOT 99  & 51509.2635 &
2.36 &   -  &   -  &   -  &   -  & 0.61 & 1.60 & 0.53 & 0.76 &   -  & HR 222\\
         & NOT 99  & 51510.3008 &
1.88 &   -  &   -  &   -  &   -  & 0.58 & 1.41 & 0.51 & 0.80 &   -  &   "   \\
         & INT 00  & 51564.1761 &
  -  &   -  &   -  &   -  &   -  & 0.58 & 1.49 & 0.49 & 0.65 & 0.58 & HR 166\\
         & INT 00  & 51566.1127 &
  -  &   -  &   -  &   -  &   -  & 0.56 & 1.58 & 0.57 & 0.65 & 0.55 &   "   \\
GJ 577   & 2.2m 99 & 51384.8466 &
0.33 & 0.22 & 0.0  & 0.0  & 0.0  & 0.0  & 0.20 & 0.23 & 0.28 & 0.28 & GJ 679\\
         & 2.2m 99 & 51386.8451 &
0.28 & 0.26 & 0.0  & 0.0  & 0.0  & 0.0  & 0.24 & 0.21 & 0.34 & 0.29 &   "   \\
         & 2.2m 99 & 51388.8314 &
0.42 & 0.22 & 0.0  & 0.0  & 0.0  & 0.0  & 0.22 & 0.19 & 0.27 & 0.27 &   "   \\
GJ 3706  & INT 00  & 51564.2814 &
  -  &   -  &   -  &   -  &   -  & 0.0  & 0.0  & 0.11 & 0.11 & 0.11 & HR 166\\
\noalign{\smallskip}
\hline
\end{tabular}
\end{flushleft}
\end{table*}

\begin{table*}
\caption[]{Absolute surface flux 
of the different chromospheric activity indicators
\label{tab:actflux}}
\begin{flushleft}
\scriptsize
\begin{tabular}{llccccccccccc}
\noalign{\smallskip}
\hline
\noalign{\smallskip}
     &      &     & \multicolumn{10}{c}{logF$_{\rm S}$ (erg cm$^{-2}$ s$^{-1}$)} \\
\cline{4-13}
\noalign{\smallskip}
Name & Obs. & MJD & \multicolumn{2}{c}{Ca II} & & & & & &
\multicolumn{3}{c}{Ca II IRT}\\
\cline{4-5}\cline{11-13}
\noalign{\smallskip}
     &      &     &
 K   & H  & H$\epsilon$ & H$\delta$ & H$\gamma$ & H$\beta$ & H$\alpha$ &
$\lambda$8498 & $\lambda$8542 & $\lambda$8662
\scriptsize
\\
\noalign{\smallskip}
\hline
\noalign{\smallskip}
{\bf LA} \\
\noalign{\smallskip}
\hline
\noalign{\smallskip}
PW And   & 2.2m 99 & 51384.1732 &
6.476 & 6.294 & 5.717 & 5.670 & 5.722 & 5.978 & 6.600 & 6.088 & 6.197 & 6.114\\
         & 2.2m 99 & 51385.0401 &
6.360 & 6.222 & 5.745 & 5.686 & 5.707 & 5.959 & 6.557 & 6.106 & 6.197 & 6.146\\
         & 2.2m 99 & 51386.1011 &
6.523 & 6.194 & 5.731 & 5.654 & 5.722 & 6.014 & 6.587 & 6.097 & 6.218 & 6.122\\
         & 2.2m 99 & 51387.0396 &
  -   &   -   &   -   & 5.686 & 5.691 & 5.950 & 6.547 & 6.070 & 6.190 & 6.138\\
         & 2.2m 99 & 51388.1258 &
6.520 & 6.140 & 5.641 & 5.619 & 5.675 & 5.987 & 6.536 & 6.070 & 6.183 & 6.114\\
         & 2.2m 99 & 51389.0818 &
6.564 & 6.451 & 5.745 &   -   &   -   & 5.978 & 6.591 & 6.088 & 6.231 & 6.122\\
         & NOT 99  & 51508.8690 &
6.612 &   -   &   -   &   -   &   -   & 6.157 & 6.656 & 6.161 & 6.290 &   -  \\
         & NOT 99  & 51509.9022 &
6.599 &   -   &   -   &   -   &   -   &   -   & 6.645 & 6.122 & 6.357 &   -  \\
V368 Cep & 2.2m 99 & 51384.0420 &
6.582 & 6.318 & 5.786 & 5.722 & 5.844 & 5.972 & 6.511 & 6.043 & 6.168 & 6.089\\
         & 2.2m 99 & 51385.0085 &
  -   & 6.298 & 5.883 & 5.638 & 5.706 & 5.956 & 6.483 & 6.043 & 6.168 & 6.089\\
         & 2.2m 99 & 51386.1725 &
6.425 & 6.256 & 5.786 & 5.722 & 5.648 & 5.886 & 6.526 & 6.018 & 6.168 & 6.089\\
         & 2.2m 99 & 51387.0935 &
6.472 & 6.270 & 5.764 & 5.793 & 5.757 & 5.956 & 6.597 & 6.055 & 6.177 & 6.110\\
         & 2.2m 99 & 51388.0347 &
6.453 & 6.298 & 5.847 & 5.395 & 5.405 & 5.972 & 6.403 & 5.992 & 6.120 & 6.078\\
         & 2.2m 99 & 51389.0296 &
6.575 & 6.377 & 5.947 &   -   &   -   & 6.187 & 6.642 & 6.078 & 6.203 & 6.203\\
         & NOT 99  & 51509.8749 &
6.514 &   -   &   -   &   -   &   -   & 5.867 & 6.337 & 5.978 & 6.300 &   -  \\
V383 Lac & 2.2m 99 & 51384.0141 &
6.699 & 6.416 & 5.887 & 5.347 & 5.530 & 5.913 & 6.412 & 6.116 & 6.230 & 6.147\\
         & 2.2m 99 & 51384.1565 &
6.658 & 6.434 & 5.927 & 5.414 & 5.421 & 5.846 & 6.426 & 6.105 & 6.238 & 6.157\\
         & 2.2m 99 & 51384.9920 &
  -   &   -   &   -   &   -   & 5.178 & 5.846 & 6.447 & 6.137 & 6.285 & 6.195\\
         & 2.2m 99 & 51386.0848 &
6.633 & 6.578 & 6.145 & 5.799 & 5.877 & 6.193 & 6.674 & 6.222 & 6.377 & 6.320\\
         & 2.2m 99 & 51387.0233 &
6.633 & 6.434 & 5.980 & 5.472 & 5.479 & 6.068 & 6.480 & 6.116 & 6.254 & 6.177\\
         & 2.2m 99 & 51388.0023 &
6.637 & 6.457 & 5.866 & 5.347 & 5.479 & 5.870 & 6.433 & 6.147 & 6.277 & 6.195\\
         & 2.2m 99 & 51388.9987 &
6.682 & 6.410 & 5.844 &   -   &   -   & 5.933 & 6.498 & 6.105 & 6.270 & 6.204\\
EP Eri   & NOT 99  & 51508.9395 &
6.331 &    -  &    -  &    -  &    -  &    -  & 5.887 & 5.749 & 6.085 &    - \\
         & NOT 99  & 51509.9946 & \\
DX Leo   & NOT 99  & 51509.2471 &
6.436 &   -   &   -   &   -   &   -   &   -   & 6.074 & 5.911 & 6.140 &   -  \\
         & NOT 99  & 51510.2689 &
6.546 &   -   &   -   &   -   &   -   &   -   & 6.056 & 5.929 & 6.117 &   -  \\
         & INT 00  & 51562.1797 &
  -   &   -   &   -   &   -   &   -   &   -   & 6.036 & 6.026 & 6.093 & 6.054\\
         & INT 00  & 51564.2121 &
  -   &   -   &   -   &   -   &   -   &   -   & 6.109 & 5.947 & 6.105 & 6.068\\
         & INT 00  & 51566.1720 &
  -   &   -   &   -   &   -   &   -   &   -   & 5.995 & 5.929 & 6.129 & 5.996\\
GJ 211   & INT 00  & 51566.0980 &
  -   &   -   &   -   &   -   &   -   &   -   & 5.232 & 5.587 & 5.763 & 5.587\\
HD 77407 & INT 00  & 51564.1974 &
   -  &   -   &   -   &   -   &   -   & 5.806 & 6.276 & 5.920 & 6.096 & 6.080\\
         & INT 00  & 51566.1601 &
   -  &   -   &   -   &   -   &   -   & 5.438 & 6.157 & 5.966 & 6.027 & 6.027\\
EK Dra   & INT 00  & 51563.3055 &
   -  &   -   &   -   &   -   &   -   & 6.356 & 6.663 & 6.270 & 6.332 & 6.342\\
         & INT 00  & 51566.3100 &
   -  &   -   &   -   &   -   &   -   & 6.337 & 6.699 & 6.235 & 6.351 & 6.322\\
\noalign{\smallskip}
\hline
\noalign{\smallskip}
{\bf UMa} \\
\noalign{\smallskip}
\hline
\noalign{\smallskip}
V834 Tau & NOT 99  & 51509.0815 &
6.655 &   -   &   -   &   -   &   -   & 5.894 & 6.535 & 6.084 & 6.295 &   -  \\
         & NOT 99  & 51510.1062 &
6.657 &   -   &   -   &   -   &   -   & 5.852 & 6.467 & 6.205 & 6.314 &   -  \\
         & INT 00  & 51566.0563 &
  -   &   -   &   -   &   -   &   -   & 5.894 & 6.486 & 5.938 & 6.076 & 5.979\\
$\pi^{1}$ UMa&NOT 99 & 51510.2862 &
  -   &   -   &   -   &   -   &   -   &   -   & 6.147 & 5.834 & 6.135 &   -  \\
         & INT 00  & 51564.1889 &
  -   &   -   &   -   &   -   &   -   &   -   & 5.868 & 5.772 & 5.862 & 5.937\\
         & INT 00  & 51566.1501 &
  -   &   -   &   -   &   -   &   -   &   -   & 5.868 & 5.772 & 5.888 & 5.913\\
GJ 503.2 & INT 00  & 51566.2634 &
  -   &   -   &   -   &   -   &   -   &   -   & 5.591 & 5.578 & 5.665 & 5.703\\
\noalign{\smallskip}
\hline
\noalign{\smallskip}
{\bf Others} \\
\noalign{\smallskip}
\hline
\noalign{\smallskip}
LQ Hya   & NOT 99  & 51509.2635 &
6.778 &   -   &   -   &   -   &   -   & 6.242 & 6.758 & 6.180 & 6.337 &   -  \\
         & NOT 99  & 51510.3008 &
6.679 &   -   &   -   &   -   &   -   & 6.220 & 6.703 & 6.164 & 6.359 &   -  \\
         & INT 00  & 51564.1761 &
  -   &   -   &   -   &   -   &   -   & 6.220 & 6.727 & 6.146 & 6.269 & 6.219\\
         & INT 00  & 51566.1127 &
  -   &   -   &   -   &   -   &   -   & 6.205 & 6.753 & 6.212 & 6.269 & 6.196\\
GJ 577   & 2.2m 99 & 51384.8466 &
6.383 & 6.206 &   -   &   -   &   -   &   -   & 6.104 & 5.979 & 6.064 & 6.064\\
         & 2.2m 99 & 51386.8451 &
6.311 & 6.279 &   -   &   -   &   -   &   -   & 6.183 & 5.939 & 6.148 & 6.079\\
         & 2.2m 99 & 51388.8314 &
6.487 & 6.206 &   -   &   -   &   -   &   -   & 6.145 & 5.896 & 6.048 & 6.048\\
GJ 3706  & INT 00  & 51564.2814 &
  -   &   -   &   -   &   -   &   -   &   -   &   -   & 5.576 & 5.576 & 5.576\\
\noalign{\smallskip}
\hline
\end{tabular}
\end{flushleft}
\end{table*}


\listofobjects


\end{document}